\documentclass[11pt]{article}
\usepackage{amsmath,amsfonts,amsthm,mathrsfs}
\newcommand{\R}{{\mathbb R}} \newcommand{\N}{{\mathbb N}}
\newcommand{\K}{{\mathbb K}} \newcommand{\Z}{{\mathbb Z}}
  \def\C{{\mathbb C}}
\newcommand{\Prm}{{\mathbb P}}
\newcommand{\wt}{\widetilde }
\newcommand{\e}{\varepsilon }
\renewcommand{\epsilon}{\varepsilon } 
\newcommand{\g}{\gamma }
 
\renewcommand{\rho}{\varrho } 
\renewcommand{\l}{{\lambda }} 
\renewcommand{\phi}{\varphi }
\newcommand{\s}{\sigma }
\renewcommand{\a}{\alpha }
\renewcommand{\b}{\beta }

\newcommand{\q}{{\rm q }}
\newcommand{\ran}{{\rm ran}}
\newcommand{\de}{{\rm det}}

\def\rs{\right>}
\def\lg{\left|}

\newtheorem{theorem}{Theorem}
\newtheorem{lemma}{Lemma}

\newtheorem{proposition}{Proposition}

\begin {document}
 \title{The Quantum Query Complexity of Elliptic PDE}

\author {Stefan Heinrich\\
Department of Computer Science\\
University of Kaiserslautern\\
D-67653 Kaiserslautern, Germany\\
e-mail: heinrich@informatik.uni-kl.de\\
homepage: http://www.uni-kl.de/AG-Heinrich}   
\date{}
\maketitle
\begin{abstract}
The complexity of the following numerical problem is studied in the quantum model of computation:
Consider a general elliptic partial differential equation of order $2m$ 
in a smooth, bounded domain $Q\subset \R^d$ with smooth coefficients and 
homogeneous boundary conditions. We seek to approximate the solution on
a smooth submanifold $M\subseteq Q$ of dimension $0\le d_1 \le d$. With the right hand side 
belonging to $C^r(Q)$, and the error being measured in the $L_\infty(M)$ norm, we prove
that the $n$-th minimal quantum error is (up to logarithmic factors) of order
$$
n^{-\min\left((r+2m)/d_1,\,r/d+1\right)}.
$$
For comparison, in the classical deterministic setting the $n$-th minimal
error is known to be of order
$
n^{-r/d},
$
for all $d_1$, while in the classical randomized setting it is (up to logarithmic factors)
$$
n^{-\min\left((r+2m)/d_1,\,r/d+1/2\right)}.
$$
\end{abstract}

\section{Introduction}
The complexity of solving elliptic problems in the classical deterministic setting was studied in 
\cite{Wer91, Wer96, Dya96, DNS05, DNS05a}. In \cite{Hei05b} such  problems were considered in the
classical randomized setting.  The quantum complexity of 
ordinary differential equation was investigated in \cite{Kac05}, while
in \cite{Kwa05} certain parabolic problems were studied in this setting. The complexity of elliptic problems 
in the quantum model of computation has not been analyzed before. This is the topic of the present paper. 
We consider a general elliptic partial 
differential equation given on a smooth domain in $\R^d$, with smooth coefficients and 
homogeneous boundary conditions. We seek to find an approximation to the solution on a given, 
$d_1$-dimensional smooth submanifold, where $0\le d_1\le d$. Thus, we consider the whole range of problems
from local solution (find the solution in a single point, $d_1=0$) up to global solution (find the full
solution, in the whole domain, $d_1=d$).
Our analysis is carried out in the quantum setting of information-based complexity theory, as developed in \cite{Hei01}.
For a study of other basic numerical problems in this framework we refer to \cite{Nov01, Hei01b, Hei04a, Hei04b,
Kac05, Kwa05, TW01, Wie04}, see also the surveys \cite{Hei03, Hei04c}. For general background 
on quantum computation we refer to the surveys  \cite{Aha98},
\cite{EHI00}, \cite{Sho00},  and the monographs \cite{Pit99}, 
\cite{Gru99},  \cite{NC00}. For the classical settings of information-based complexity theory we refer to 
\cite{TWW88,Nov88, Hei93}.

This paper can be considered as  a continuation of \cite{Hei05a, Hei05b}. The approximation of weakly singular 
integral operators plays a key role again. In some situations,  techniques from \cite{Hei05a, Hei05b}
can also be applied to the quantum setting, while in others entirely different approaches 
are needed. In particular, a number of new tools for the general
quantum setting of information-based complexity has to be developed.

The paper is organized as follows. In section 2 we describe the quantum setting, general results about quantum $n$-th minimal 
errors are derived in section 3. In section 4 we study weighted mean computation and
integration. These are preparations for section 5, in which we are concerned with quantum approximation of weakly singular operators.
Section 6 contains the statement and the proof of the main result about the query complexity of elliptic PDE. 
Finally, in section 7 we recall the respective
results of the classical deterministic and randomized settings and compare them with the quantum setting.

\section{Notation}

A numerical problem is given by a tuple $\mathcal{P}=(F,G,S,K,\Lambda)$, where
$F$ is a non-empty set, $G$ a normed space over $\K$, where $\K$ 
stands for the set of real or complex numbers, $S$ a mapping from $F$ to $G$, $K$ a non-empty
set and $\Lambda$ a non-empty set of mappings from $F$ to $K$.   We seek to approximate $S(f)$ for $f\in F$ 
by means of quantum computations. 

Usually $F$ is a set in a function space, $S$ is the solution
operator, which maps the input $f\in F$ to the exact solution $S(f)$, and we want to approximate $S(f)$.
$\Lambda$ is usually a set of linear functionals, supplying information $\lambda(f)$ about $f$ through which the algorithm 
can access the input $f$. $K$ is mostly $\R$ or $\C$, $G$ is a space 
containing both the solutions and the approximations,  and the error is measured in the norm of $G$. 

In the sequel it will be convenient to consider $f\in F$ also as a function on $\Lambda$ with values in $K$
by setting $f(\lambda):=\lambda(f)$. Let $\mathcal{F}(\Lambda,K)$ denote the set of all functions from $\Lambda$ to $K$. 

Let $H_1$ be the 
two-dimensional complex Hilbert space $\C^2$, with its unit vector
basis $\{e_0,e_1\}$, let
$$
 H_m=\underbrace{H_1\otimes\dots\otimes H_1}_{m}, 
$$
equipped with the tensor
Hilbert space structure. 
Denote
$$\Z[0,N) := \{0,\dots,N-1\}$$
for $N\in\N$ (we  write $\N= \{1,2,\dots \}$ 
and $\N_0=\N\cup\{0\}$).
Let $\mathcal{C}_m = \{\lg i\rs:\, i\in\Z[0,2^m)\}$ be the canonical basis of
$H_m$, where  $\lg i \rs$ stands for 
$e_{j_0}\otimes\dots\otimes e_{j_{m-1}}$ with $i=\sum_{k=0}^{m-1}j_k2^{m-1-k}$.
 Let $\mathcal{U}(H_m)$ denote the set of unitary operators on $H_m$. 

A quantum query  on $F$ is given by a tuple
\begin{equation*}
Q=(m,m',m'',Z,\tau,\beta),
\end{equation*}
where $m,m',m''\in \N, m'+m''\le m, Z\subseteq \Z[0,2^{m'})$ is a nonempty 
subset, and
$$\tau:Z\to \Lambda$$
$$\beta:K\to\Z[0,2^{m''})$$
are arbitrary mappings. Let $m(Q):=m$ denote the number of qubits of $Q$. 

Given a query $Q$, we define for each $f\in F$ the unitary operator 
$Q_f\in\mathcal{U}(H_m)$ by setting for  
$\lg i\rs\lg x\rs\lg y\rs\in \mathcal{C}_m
=\mathcal{C}_{m'}\otimes\mathcal{C}_{m''}\otimes\mathcal{C}_{m-m'-m''}$:
\begin{equation*}
Q_f\lg i\rs\lg x\rs\lg y\rs=
\left\{\begin{array}{ll}
\lg i\rs\lg x\oplus\beta(f(\tau(i)))\rs\lg y\rs &\quad \mbox {if} \quad i\in Z\\
\lg i\rs\lg x\rs\lg y\rs & \quad\mbox{otherwise,} 
 \end{array}
\right. 
\end{equation*}
where $\oplus$ means addition modulo $2^{m''}$. 

A quantum algorithm on $F$  with no measurement is a tuple
\begin{equation*}
A=(Q,(U_j)_{j=0}^n),
\end{equation*}
where $Q$ is a quantum query on $F$, $n\in\N_0$  and
$U_{j}\in \mathcal{U}(H_m)\,(j=0,\dots,n)$, with $m=m(Q)$.
Given $f\in F$,
we define $A_f\in \mathcal{U}(H_m)$ as
\begin{equation*}
A_f = U_n Q_f U_{n-1}\dots U_1 Q_f U_0.
\end{equation*}
We denote by $n_q(A):=n$ the number of queries and by $m(A)=m=m(Q)$ the 
number of qubits of $A$. Let $(A_f(x,y))_{x,y\in \Z[0,2^m)}$ 
be the matrix of the 
transformation $A_f$ in the canonical basis $\mathcal{C}_{m}$.

A quantum algorithm from $F$ to $G$ with $k$ measurements  is a tuple
$$
A=((A_l)_{l=0}^{k-1},(b_l)_{l=0}^{k-1},\varphi),
$$ 
where $k\in\N$, $A_l\;(l=0,\dots,k-1)$ are quantum algorithms
on $F$ with no measurements, 
$$
b_0\in\Z[0,2^{m_0}), 
$$
$$
b_l:\prod_{i=0}^{l-1}\Z[0,2^{m_i}) \to \Z[0,2^{m_l})\quad 
(1\le l \le k-1),
$$
where $m_l:=m(A_l)$, and 
$$
\varphi:\prod_{l=0}^{k-1}\Z[0,2^{m_l}) \to G.
$$
The output of $A$ at input $f\in F$ will be a probability measure $A(f)$ on $G$, 
defined as follows: First put
\begin{eqnarray}   
p_{A,f}(x_0,\dots, x_{k-1})&=&
|A_{0,f}(x_0,b_0)|^2 |A_{1,f}(x_1,b_1(x_0))|^2\dots\nonumber\\
&&\dots |A_{k-1,f}(x_{k-1},b_{k-1}(x_0,\dots,x_{k-2}))|^2.\nonumber
\end{eqnarray}
Then define $A(f)$ by setting for any subset $C\subseteq G$
\begin{equation*}
A(f)(C)=\sum_{\phi(x_0,\dots,x_{k-1})\in C}p_{A,f}(x_0,\dots, x_{k-1}).
\end{equation*}
Let 
$n_q(A):=\sum_{l=0}^{k-1} n_q(A_l)$
denote the number of queries used by $A$.
For more details and background see \cite{Hei01}.
Below we  use the term `quantum algorithm', 
meaning a quantum algorithm with measurement(s).
Note that a quantum query on $F$ (respectively, a quantum algorithm from $F$ to $G$) can also be considered as a
quantum query on $\mathcal{F}(\Lambda,K)$  (respectively, a quantum algorithm from $\mathcal{F}(\Lambda,K)$ to $G$),
and vice versa.

The above definition 
simplifies essentially for an algorithm with one measurement, which is given by
$$
A=(A_0,b_0,\phi),\quad A_0=(Q,(U_j)_{j=0}^n).
$$
The quantum computation is carried out on $m:=m(Q)$ qubits. For $f\in F$ the 
algorithm starts in the state  $\lg b_0 \rs$ and produces
$$
\lg \psi_f \rs=U_n Q_f U_{n-1}\dots U_1 Q_f U_0\lg b_0 \rs.
$$
Let
$$ 
\lg \psi_f \rs=\sum_{i=0}^{2^m-1}a_{i,f}\lg i \rs.
$$
 Then the output takes the 
value $\phi(i)\in G$ with probability $|a_{i,f}|^2$. As shown in \cite{Hei01},
Lemma 1, for each algorithm $A$ with $k$ measurements there is an algorithm 
$\wt{A}$ with one measurement such that $A(f)=\wt{A}(f)$ for all $f\in F$ and 
$\wt{A}$ uses just twice the number of queries of $A$.

For $\theta\ge 0$ and a quantum algorithm $A$ we 
define the (probabilistic) error at $f\in F$ as follows. Let $\zeta$ be a random variable with
distribution $A(f)$. Then
\begin{equation*}
e(S,A,f,\theta)=\inf\left\{\varepsilon\ge0\,\,|\,\,\Prm\{\|S(f)-\zeta\|>\varepsilon\}\le\theta
\right\}
\end{equation*}
(observe that this infimum is always attained). Let
$$
e(S,A,F,\theta)=\sup_{f\in F} e(S,A,f,\theta) 
$$
(this quantity can take the value $+\infty$). Furthermore, we set
\begin{eqnarray*}
\lefteqn{e_n^\q(S,F,\theta)}\\
&=&\inf\{e(S,A,F,\theta)\,\,|\,\,A\,\,
\mbox{is any quantum algorithm with}\,\, n_q(A)\le n\}.
\end{eqnarray*}
 We denote
$$
e(S,A,f)=e(S,A,f,1/4)
$$
and similarly,
$$
e(S,A,F)=e(S,A,F,1/4),\quad e_n^\q(S,F)=e_n^\q(S,F,1/4).
$$
The quantity $e_n^\q(S,F)$ is the 
$n$-th minimal query error, 
that is, the smallest error which can be reached using at most $n$ queries.
Note that it essentially suffices to study
$e_n^\q(S,F)$ instead of $e_n^\q(S,F,\theta)$, 
since with $\mathcal{O}(\nu)$ repetitions, the error probability 
can be reduced to $2^{-\nu}$ (see Lemmas 3, 4 and Corollary 1 of \cite{Hei04a}).
 
The quantum query complexity is defined for $\varepsilon > 0$ by 
\begin{eqnarray*}
\lefteqn{\mbox{comp}_\varepsilon^\q(S,F)=}\\
&&\min\{n_q(A)\,\,|\,\, A\,\,\mbox{is any quantum 
algorithm with}\,\, e(S,A,F) \le \varepsilon\}
\end{eqnarray*}
(we put $\mbox{comp}_\varepsilon^\q(S,F)=+\infty$ if there is no such algorithm).
It is easily checked that these functions are inverse to each other in the 
following sense: For all $n\in \N_0$ and $\varepsilon > 0$,
$e_n^\q(S,F)\le \varepsilon$ if and only if
$\mbox{comp}_{\varepsilon_1}^\q(S,F)\le n$ for all $\varepsilon_1 > \varepsilon$.
Hence it suffices to determine one of them. We shall principally choose the first one.

Note that the definition of a numerical problem we presented here corresponds to that used in \cite{Hei05a,Hei05b} 
for the classical settings, and
is slightly more general than the one in previous papers on the quantum setting
\cite{Hei01,Hei01b,Hei04a,Hei04b}. There $F$ was always a set of functions on some set $D$. We get back to this
setting by considering, as done above, each $f$ as a function on $\Lambda$ and defining $D=\Lambda$. (Such an approach
has already been outlined at the end of \cite{Hei01}.) The mapping that sends $f\in F$ to the corresponding function
$(f(\l))_{\l\in \Lambda}$ needs not to be one-to-one, in general. 
Nevertheless, all general results of \cite{Hei01,Hei01b,Hei04a,Hei04b} carry over in an obvious way, with literally identical proofs.

\section{Some general results}

Let $\wt{\mathcal{P}}=(\wt{F},\wt{G},\wt{S},\wt{K},\wt{\Lambda})$ be another numerical problem.
Suppose we have an algorithm for problem $\wt{\mathcal{P}}$, and we want to construct one for 
problem $\mathcal{P}$. Furthermore,  for each input $f\in F$ of problem $\mathcal{P}$ we can produce an
input $R(f)$ for problem $\wt{\mathcal{P}}$ such that  $S(f)=\Psi\circ\wt{S}\circ R(f)$ with a certain mapping $\Psi: \wt{G}\to G$.
Finally, each information about $R(f)$ can be obtained from
$\kappa$ suitable informations about $f$.  
Then we say that problem $\mathcal{P}$ reduces to $\wt{\mathcal{P}}$. Let us specify the assumptions. 

Let 
$R:F\to\wt{F}$ be a mapping such that there exist a $\kappa \in \N$, mappings 
$\eta_j:\wt{\Lambda}\to\Lambda \quad (j=0, \dots, \kappa-1)$
and $\rho:\wt{\Lambda}\times K^\kappa\to \wt{K}$ with
\begin{equation}
\label{XG3}
(R(f))(\wt{\lambda})=\rho(\wt{\lambda}, f(\eta_0(\wt{\lambda})), \dots,f(\eta_{\kappa-1}(\wt{\lambda})))
\end{equation}
for all $f\in F$ and $\wt{\lambda}\in\wt{\Lambda}$.
Furthermore, let   
$\Psi:\wt{G}\to G$ be a Lipschitz mapping and assume that 
\begin{equation}
\label{XG4}
S=\Psi\circ\wt{S}\circ R. 
\end{equation}
Note that (\ref{XG3}) defines also a mapping 
$$
R:\mathcal{F}(\Lambda,K) \to \mathcal{F}(\wt{\Lambda},\wt{K})
$$
(we use the same notation $R$), where  $\mathcal{F}(\Lambda,K)$ stands for the set of all mappings from $\Lambda$ to $K$. 

\begin{lemma}\label{lem:1} 
Let $F_0\subseteq F$ be any nonempty subset.
Suppose that for each $\delta>0$ and each finite subset $\Lambda_0\subseteq \Lambda$ there are mappings
$$
\theta:K\to K,\quad \Theta:F_0\to F
$$ 
such that $\theta(K)$ is a finite set,
\begin{equation}
\label{U4}
(\Theta(f))(\l)=\theta(f(\l))\quad(f\in F_0,\, \l\in \Lambda_0),
\end{equation}
and
\begin{equation}
\label{U2}
\sup_{f\in F_0}\|S(f)-S(\Theta(f))\|\le \delta.
\end{equation}
Then for all $n\in \N_0$, 
\begin{equation}
\label{U3}
e_{2\kappa n}^\q(S,F_0)\le \|\Psi\|_{\rm Lip}\, e_n^\q(\wt{S},\wt{F}).
\end{equation}
\end{lemma}
\begin{proof} 
Let $\delta>0$, $n\in\N_0$ and let $\wt{A}$ be any quantum algorithm from $\wt{F}$ to $\wt{G}$ with $n_q(\wt{A})\le n$
and 
$$
e(\wt{S},\wt{A},\wt{F})\le e_n^\q(\wt{S},\wt{F})+\delta.
$$
Let
$$
\wt{A}=((\wt{A}_l)_{l=0}^{k-1},(\wt{b}_l)_{l=0}^{k-1},\wt{\phi}),\quad 
\wt{A}_l=(\wt{Q}_l,(\wt{U}_{l,j})_{j=0}^{n_l}),
$$
$$
\wt{Q}_l=(\wt{m}_l,\wt{m}_l',\wt{m}''_l,\wt{Z}_l,\wt{\tau}_l,\wt{\beta}_l),
$$
where  $\wt{Z}_l\subseteq \Z[0,2^{\wt{m}_l'})$  and
$$
\wt{\tau}_l:\wt{Z}_l\to \wt{\Lambda}, \quad
\wt{\beta}_l:\wt{K}\to\Z[0,2^{\wt{m}''_l}).
$$
Denote 
$$
\wt{\Lambda}_0=\{\wt{\tau}_l(i)\,:\, i\in \wt{Z}_l,\, l=0,\dots, k-1\}
$$
and
$$
\Lambda_0=\{\eta_j(\wt{\l})\,:\,\wt{\l}\in \wt{\Lambda}_0\}.
$$
Now let $\theta$ and $\Theta$ be according to the assumptions, and choose $m^{*}$ so
that $|\theta(K)|\le 2^{m^{*}}$. It is easily checked that one can find 
$$
\beta: K\to \Z[0,2^{m^{*}})
$$
and 
$$
\g:\Z[0,2^{m^{*}})\to K
$$
such that $\g\circ \beta=\theta$.
 Define 
$$
\bar{\rho}:\wt{\Lambda}\times \Z[0,2^{m^{*}})^\kappa \to \wt{K}
$$
for $\wt{\l}\in \wt{\Lambda}$, $t_0,\dots,t_{\kappa-1}\in \Z[0,2^{m^{*}})^\kappa$ by 
$$
\bar{\rho}(\wt{\l},t_0,\dots,t_{\kappa-1})=\rho(\wt{\l},\gamma(t_0),\dots,\gamma(t_{\kappa-1}))
$$
and 
$$
\bar{R}:\mathcal{F}(\Lambda,K) \to \mathcal{F}(\wt{\Lambda},\wt{K})
$$ 
by 
$$
\bar{R}(f) = R(\theta\circ f).
$$
Then for $f\in \mathcal{F}(\Lambda,K)$ and $\wt{\l}\in \wt{\Lambda}$,
\begin{eqnarray*}
(\bar{R}(f))(\wt{\l})&=&(R(\theta\circ f))(\wt{\l})
=\rho(\wt{\l}, \theta\circ f(\eta_0(\wt{\l})),\dots,\theta\circ f(\eta_{\kappa-1}(\wt{\l})))\\
&=&\bar{\rho}(\wt{\l}, \beta\circ f(\eta_0(\wt{\l})),\dots,\beta\circ f(\eta_{\kappa-1}(\wt{\l}))).
\end{eqnarray*}
 Thus, the mapping $\bar{R}$ is of the form 
needed to apply Corollary 1 of \cite{Hei01b}. Accordingly, considering  $\wt{A}$ as a quantum algorithm from
$\mathcal{F}(\wt{\Lambda},\wt{K})$ to $\wt{G}$, we can find a quantum algorithm 
$A$ from $\mathcal{F}(\Lambda,K)$ to $\wt{G}$ with
$n_q(A)=2\kappa n_q(\wt{A})$ and 
$$
A(f)=\wt{A}(\bar{R}(f)) \quad (f\in \mathcal{F}(\Lambda,K)).
$$
For $\wt{\l}\in \wt{\Lambda}_0$ we have $\eta_j(\wt{\l})\in \Lambda_0$, and therefore, by assumption (\ref{U4}), for $f\in F_0$, 
\begin{eqnarray*}
(\bar{R}(f))(\wt{\l})&=&
\rho(\wt{\l}, \theta\circ f(\eta_0(\wt{\l})),\dots,\theta\circ f(\eta_{\kappa-1}(\wt{\l})))\\
&=&\rho(\wt{\l}, (\Theta(f))(\eta_0(\wt{\l})),\dots,(\Theta(f))(\eta_{\kappa-1}(\wt{\l})))\\
&=& (R(\Theta(f)))(\wt{\l}).
\end{eqnarray*}
This implies 
\begin{equation}
\label{U5}
\wt{Q}_{l,\bar{R}(f)}=\wt{Q}_{l,R(\Theta(f))}   \quad (l=0,\dots,,k-1),
\end{equation}
and consequently
\begin{equation}
\label{U6}
A(f)=\wt{A}(\bar{R}(f))=\wt{A}(R(\Theta(f)))\quad (f\in F_0).
\end{equation}
Now fix $f\in F_0$ and let $\zeta$ be a random variable with distribution $A(f)$. We have, by assumption (\ref{U2})
\begin{eqnarray}
\label{U7}
\|S(f)-\Psi(\zeta)\|&\le& \|S(f)-S(\Theta(f))\|+\|S(\Theta(f))-\Psi(\zeta)\|\nonumber\\
&\le& \|S(\Theta(f))-\Psi(\zeta)\|+\delta.
\end{eqnarray}
Furthermore, by (\ref{XG4}), 
\begin{eqnarray}
\label{QC3}
\|S(\Theta(f))-\Psi(\zeta)\|&=&\|\Psi\circ \wt{S}\circ R(\Theta(f))-\Psi(\zeta)\|\nonumber\\
&\le& \|\Psi\|_{\rm Lip}\,\|\wt{S}\circ R(\Theta(f))-\zeta\|.
\end{eqnarray}
Since $\Theta(f)\in F$, we have $R(\Theta(f))\in \wt{F}$. Moreover, by (\ref{U6}), the distribution of $\zeta$ is equal to 
$\wt{A}(R(\Theta(f)))$. Therefore we get with probability at least 3/4,
$$
\|\wt{S}\circ R(\Theta(f))-\zeta\|\le e(\wt{S},\wt{A},\wt{F}),
$$
and hence, by (\ref{QC3}),
\begin{eqnarray*}
\|S(\Theta(f))-\Psi (\zeta)\|&\le& \|\Psi\|_{\rm Lip}\, e(\wt{S},\wt{A},\wt{F})\\
&\le& \|\Psi\|_{\rm Lip}\,(e_n^\q(\wt{S},\wt{F})+\delta).
\end{eqnarray*}
$\Psi (\zeta)$ is a random variable with distribution $\Psi(A)(f)$ --  the output of the quantum algorithm $\Psi(A)$ 
from $F$ to $G$ (compare Lemma 2 of \cite{Hei01} and the definition before it), an algorithm with not more than $2\kappa n$ queries.
 This implies (\ref{U3}).
\end{proof}

We need some further notation. For a linear space $X$ we denote by $X^{\#}$ the algebraic dual, that is, the  space of all
linear (not necessarily continuous) functionals on X, and by $X^*$ the dual space, which is the space of all continuous linear 
functionals on $X$.
 Given a subset $F_0$ of a normed space $X$ and $\delta>0$, we denote by
$F_0^\delta$ the closed $\delta$-neighbourhood of $F_0$, that is, the set
$$
F_0^\delta=\cup_{x\in F_0} B(x,\delta),
$$
with $B(x,\delta)$ being the closed ball of radius $\delta$ around $x$. The unit ball
$B(0,1)$ of $X$ is denoted by $B_X$.

\begin{lemma}
\label{lem:2}
Let $K=\K$, let $F$ be a bounded subset of a normed space $X$, and let $\emptyset \ne F_0\subseteq F$. Assume that either\\
(i) there is a $\delta_0>0$ such that $F_0^{\delta_0}\subseteq F$ or\\
(ii) $F$ is a non-zero multiple of the unit ball of $X$.\\
Furthermore, let $\Lambda_0\subset X^{\#}$ be a finite, linearly independent
set with 
$$
sup_{f\in F_0}|f(\l)|<\infty \quad (\l\in \Lambda_0).
$$
Then for each $\delta>0$ there are mappings 
$$
\theta:\K\to \K,\quad \Theta:F_0\to F
$$ 
such that $\theta(\K)$ is a finite set,
\begin{equation}
\label{RU4}
(\Theta(f))(\l)=\theta(f(\l))\quad(f\in F_0,\, \l\in \Lambda_0),
\end{equation}
and
\begin{equation}
\label{RU2}
\sup_{f\in F_0}\|f-\Theta(f)\|\le \delta.
\end{equation}
\end{lemma}
\begin{proof} We can assume $\delta\le \delta_0<1$.
 The linear independence of $\Lambda_0$ implies
that for each $\l\in\Lambda_0$ there is a $g_\l\in X$ with 
$g_\l(\l)=1$ and $g_\l(\mu)=0$ for $\mu\in \Lambda_0\setminus\{\l\}$. 
Define
\begin{equation}
\label{U12}
M_1=\max_{\l\in\Lambda_0}\|g_\l\|,\quad M_2=\sup_{\l\in\Lambda_0, f\in F_0}|f(\l)|,\quad M_3=\sup_{f\in F_0} \|f\|,
\end{equation}
\begin{equation}
\label{U14}
\delta_1=\delta/(M_3+1),
\end{equation}
and choose any $\theta_0: \K\to \K$ such that $\theta_0(\K)$ is finite and
\begin{equation}
\label{U11}
|a-\theta_0(a)|\le M_1^{-1}|\Lambda_0|^{-1}\delta_1\min(M_3,1)\quad(|a|\le M_2). 
\end{equation}
Now we define $\theta:\K\to\K$ by setting for $a\in\K$,
$$
\theta(a)=\theta_0((1-\delta_1)a),
$$
and $\Theta:F_0\to X$ by
\begin{equation}
\label{U10}
\Theta(f)=(1-\delta_1)f-\sum_{\l\in\Lambda_0}((1-\delta_1)f(\l)-\theta_0((1-\delta_1)f(\l))g_\l.
\end{equation}
Then for $f\in F_0$, $\mu\in \Lambda_0$,
\begin{eqnarray*}
\lefteqn{ (\Theta(f))(\mu)}\\
&=&(1-\delta_1)f(\mu)-\sum_{\l\in\Lambda_0}((1-\delta_1)f(\l)-\theta_0((1-\delta_1)f(\l))g_\l(\mu)\\
&=&\theta_0((1-\delta_1)f(\mu))=\theta(f(\mu)),
\end{eqnarray*}
which verifies (\ref{RU4}). Moreover, we have, by (\ref{U12}) and (\ref{U11}),
\begin{eqnarray}
\label{U13}
\lefteqn{\left\|\sum_{\l\in\Lambda_0}((1-\delta_1)f(\l)-\theta_0((1-\delta_1)f(\l))g_\l \right\|  }\nonumber\\
&\le&\left\|\sum_{\l\in\Lambda_0}|(1-\delta_1)f(\l)-\theta_0((1-\delta_1)f(\l))| g_\l \right\|\le  \min(M_3,1)\delta_1.
\end{eqnarray}
Hence, by (\ref{U10}), (\ref{U13}), and (\ref{U14})
$$
\|f-\Theta(f)\|\le \delta_1\|f\|+\delta_1 \le (M_3+1) \delta_1=\delta\le \delta_0,
$$
which proves (\ref{RU2}). Furthermore, it shows that in case of condition (i), $\Theta(f)\in F$ for all $f\in F_0$.
If condition (ii) is fulfilled, that is, $F=a_0B_X$ for some $a_0>0$, we argue as follows:
$$
\|\Theta(f)\|\le\|(1-\delta_1)f\|+M_3\delta_1\le (1-\delta_1)a_0+\delta_1 a_0=a_0,
$$
thus, again, $\Theta(f)\in F$ for all $f\in F_0$.
\end{proof}
\begin{proposition} 
\label{pro:1} 
Let $K=\K$. Assume that $S,\wt{S},R,\Psi$ are as above (\ref{XG3}), (\ref{XG4}), that $F$ is a bounded subset of a normed space $X$, 
and $\Lambda$ is a linearly independent subset of  $X^\#$. Let $F_0$ be a nonempty subset of $F$ and assume that either\\
(i) $F_0^{\delta_0}\subseteq F$ for some $\delta_0>0$, or\\ 
(ii) $F$ is a non-zero multiple of the unit ball of $X$. \\
Furthermore suppose $\sup_{f\in F_0}|f(\l)|<\infty$ for each $\l\in \Lambda$ and
 $S$ is uniformly continuous on $F$. 
Then for all $n\in \N_0$, 
\begin{equation*}
e_{2\kappa n}^\q(S,F_0)\le \|\Psi\|_{\rm Lip}\, e_n^\q(\wt{S},\wt{F}).
\end{equation*}
\end{proposition}
\begin{proof}
This is a direct consequence of Lemmas \ref{lem:1}, \ref{lem:2}, and the uniform continuity of $S$.
\end{proof}
In previous papers on quantum complexity \cite{Hei01, Hei01b, Hei04b} the analysis of reductions was somewhat
cumbersome, since a certain discretization had to be applied in each particular case. Proposition \ref{pro:1} simplifies the analysis
and will be used for a number of reductions, in particular in sections 5 and 6.

Next we recall additivity properties of the quantum minimal error, see \cite{Hei01b}, Corollary 2.
\begin{proposition}
\label{pro:3}
 Let  $p\in\N$
and let $S_l:F\to G$ $(l=1,\dots,p)$ be mappings such that 
$S(f)=\sum_{l=1}^p S_l(f)\quad(f\in F)$. 
 Let 
$\nu_1,\dots,\nu_p\in \N$ be numbers satisfying
$$
\sum_{l=1}^p e^{-\nu_l/8}\le \frac{1}{4}. $$ 
Then for all $n_1,\dots,n_p\in\N_0$ 
$$
e_{\sum_{l=1}^p \nu_l n_l}^\q(S,F)\le2\sum_{l=1}^p e_{n_l}^\q(S_l,F).
$$
\end{proposition}
Given a subset $B\subseteq X$
of a normed space $X$, we denote by $\mathscr{C}(B)$ the set of all precompact subsets of $B$. A set $H\subset X^\#$ is called 
linearly independent over a non-empty set $B\subseteq X$, if the restrictions of elements of $H$ to ${\rm span}(B)$
form a linearly independent subset of $({\rm span}(B))^\#$.  

Finally we state multiplicativity properties of the minimal quantum error.  
\begin{proposition}\label{pro:4} Let $K=\K$. Assume that $F$ is a subset of a normed space $Y$, 
that $\Lambda$ is a linearly independent subset of $Y^\#$ and $\sup_{f\in F}|f(\l)|<\infty$ for each $\l\in \Lambda$.
Let $J:F\to Y$ be the embedding map, let 
$T:Y\to G$ be a bounded linear operator and assume that $S=TJ$.
Furthermore,
let $\nu_1,\nu_2\in \N$ be any numbers with
\begin{equation}
\label{KB1}
e^{-\nu_1/8}+e^{-\nu_2/8}-e^{-(\nu_1+\nu_2)/8}\le 1/4.
\end{equation}
Along with 
$\mathcal{P}=(F,G,S,\K,\Lambda)$ we consider the problems $(F,Y,J,\K,\Lambda)$ and $(B_Y,G,T,\K,\Lambda)$.
 Then
for all $n_1, n_2\in \N_0$, 
\begin{equation}
\label{AJ4}
e^\q_{\nu_1 n_1+2\nu_2 n_2}(S,F)
\le 
4 e^\q_{n_1}(J,F)\,e^\q_{n_2}(T,B_Y).
\end{equation}
If, moreover, $F$ is a precompact subset of $Y$ and $\Lambda$ is linearly independent over $F$, then
\begin{equation}
\label{QT3}
e^\q_{\nu_1 n_1+2\nu_2 n_2}(S,F)
\le 
4e^\q_{n_1}(J,F)\sup_{\mathcal{E}\in\mathscr{C}(B_Y)} e^\q_{n_2}(T, \mathcal{E}).
\end{equation}
\end{proposition}

The first part, relation (\ref{AJ4}), was proved in \cite{Hei04a}, Proposition 1 and 
Corollary 3. As already mentioned at the end of the previous section, this result was formulated for a slightly 
less general type of numerical problem, but the proof of (\ref{AJ4}) is literally the same as that of Proposition 1 and 
Corollary 3 in \cite{Hei04a}. 

The specific form of multiplicativity stated in (\ref{QT3}) (a single $e_n^\q(S,F)$ is replaced by a supremum 
over a family of subsets of $F$) will be needed in sections 5 and 6. For further explanation we refer to the remark after 
Proposition \ref{pro:Q2}.\medskip

\noindent {\it Proof of Proposition \ref{pro:4}.}
It remains to prove the second part, relation (\ref{QT3}). We  derive it from the first part, (\ref{AJ4}).
Denote 
\begin{equation}
\label{KC4}
e^\q_{n_1}(J,F)=\s
\end{equation}
and fix any $\delta>0$. Let $A=((A_l)_{l=0}^{k-1},(b_l)_{l=0}^{k-1},\varphi)$ 
be a quantum algorithm from $F$ to $Y$ with $n_q(A)\le n_1$ and 
\begin{equation}
\label{QT4}
e(J,A,F)\le \s+\delta.
\end{equation}
Let $\zeta$ be a random variable with distribution $A(f)$. 
Observe that, by definition (see section 2), $\zeta$ takes values in the finite set 
$$
Y_0=\phi\left(\prod_{l=0}^{k-1}\Z[0,2^{m_l})\right)\subset Y.
$$
Define  $\mathcal{E}_0\subset Y$ to be the closed, absolutely convex hull of $F\cup Y_0$, 
and put 
$$
\mathcal{E}=B_Y \cap \frac{2}{\s+\delta}\mathcal{E}_0.
$$
Since $F\cup Y_0$ is  precompact in $Y$, so are $\mathcal{E}_0$ and $\mathcal{E}$. 
Moreover, there is a $\g>0$ such that $F\cup Y_0\subseteq \g B_Y$. Hence, 
\begin{equation}
\label{KC3}
F\cup Y_0\subseteq \max\left(\g,\frac{\s+\delta}{2}\right)\mathcal{E}.
\end{equation}
For any $f\in F$ we have 
$$
f-\zeta\in 2\mathcal{E}_0
$$
and by (\ref{QT4}), with probability at least 3/4,
$$
f-\zeta\in (\s+\delta) B_Y.
$$
Consequently, with probability at least 3/4,
\begin{equation}
\label{QT5}
f-\zeta\in (\s+\delta) \mathcal{E}.
\end{equation}
Since $\mathcal{E}$ is  a closed, absolutely convex, and bounded subset of $Y$, it defines a norm $\|\,.\,\|_{\mathcal{E}}$
on $E={\rm span}(\mathcal{E})$ as follows
$$
\|\,y\,\|_{\mathcal{E}}=\inf\{\theta>0\,:\, y\in \theta \mathcal{E}\}\quad (y\in E),
$$
and $\mathcal{E}$ is the unit ball of $(E,\|\;\|_{\mathcal{E}})$. By (\ref{KC3}), $F\subset E$ and $Y_0\subset E$. 
Define $J_E: F\to E$ and $\phi_E$ to be $J$ and $\phi$, respectively, considered as
mappings into $E$. 
Define $A_E= ((A_l)_{l=0}^{k-1},(b_l)_{l=0}^{k-1},\phi_E)$. Then $A_E$ is a
quantum algorithm from $F$ to $E$ with $n_q(A_E)\le n_1$. By (\ref{QT5}) and (\ref{KC4}),
\begin{equation*}
e_{n_1}^\q(J_E,F)\le e(J_E,A_E,F)\le \s+\delta=e_{n_1}^\q(J,F)+\delta.
\end{equation*}
Note that, since $F\subseteq E$, $\Lambda$ is linearly independent over $E$. Furthermore, since $B_E=\mathcal{E}$ is bounded in
$Y$, the restriction of $T$ to $E$ is a bounded linear operator from $E$ to $G$. Applying now the first part of Proposition \ref{pro:4},
 we get 
\begin{equation*}
e^\q_{\nu_1 n_1+2\nu_2 n_2}(S,F)\le 
4 e^\q_{n_1}(J_E,F)\,e^\q_{n_2}(T,B_E)\le 4 (e^\q_{n_1}(J,F)+\delta)\,e^\q_{n_2}(T,\mathcal{E}),
\end{equation*}
which gives the desired result, since $\delta>0$ was arbitrary.
\qed\medskip

Let $e_n^\de(S,F)$ denote the (classical) $n$-th minimal deterministic error, that is,  the minimal error among all
deterministic, adaptive algorithms using at most $n$ informations (see, e.g., \cite{Hei05a}, section 4).
We want to apply relations (\ref{AJ4}) and (\ref{QT3}) with $e^\q_{n_1}(J,F)$ replaced by 
$e^\de_{n_1}(J,F)$. An estimate of $e_n^\q(S,F)$ by $e_n^\de(S,F)$ is not obvious, since classical deterministic algorithms 
can use information
with values in $\K$ directly, while quantum algorithms can use them only through a finite encoding. We therefore supply the following
\begin{lemma} 
\label{lem:3}
Let $K=\K$,
assume that $F$  is the unit ball of a normed space $X$, $\Lambda \subseteq X^*$, and $S$ is a bounded linear 
operator from 
$X$ to $G$.
Then for all $n\in \N_0$
\begin{equation}
\label{VU1}
e^\q_n(S,F)\le 2 e^\de_n(S,F).
\end{equation}
\end{lemma}
\begin{proof}
Let $\delta>0$. It is well-known (see \cite{TWW88}, Theorem  5.2.1 and Corollary 5.2.1) that there is a 
nonadaptive deterministic algorithm $\wt{A}$ using at most $n$ informations
such that 
\begin{equation}
\label{RU5}
 \sup_{f\in F} \|S(f)-\wt{A}(f)\|\le 2 e^\de_n(S,F)+\delta.
\end{equation}
Such an $\wt{A}$ has the following form: There are $\l_0,\dots,\l_{n-1}\in \Lambda$ and a mapping $\phi:\K^n\to G$ such that
\begin{equation}
\label{RU6}
\wt{A}(f)=\phi(f(\l_0),\dots, f(\l_{n-1}))\quad (f\in F).
\end{equation}
Without loss of generality we may assume that $\Lambda_0=\{\l_0,\dots,\l_{n-1}\}\subset X^*$ is a linearly independent set
(if not, we pass to an independent subset and omit the rest by suitably modifying $\phi$).
Let $\theta$ and $\Theta$ be the mappings which result from the application of Lemma \ref{lem:2}, case (ii). Put $m'=1$, 
choose $m''\in\N$ such that $|\theta(\K)|\le 2^{m''}$ and let $m=m'+m''$. We represent 
(as done before, in the proof of Lemma \ref{lem:1}) 
$$
\theta=\g\circ\b
$$
with $\b:\K\to \Z[0,2^{m''})$ and $\g:\Z[0,2^{m''})\to \K$. Furthermore, we identify $\Z[0,2^m)$ with $\{0,1\}\times\Z[0,2^{m''})$.
Define $\bar{\phi}:(\Z[0,2^m))^n\to G$ for 
$$
(b_i,a_i)\in \{0,1\}\times\Z[0,2^{m''})=\Z[0,2^m)\quad(i=0,\dots,n-1)
$$
by setting
$$
\bar{\phi}((b_0,a_0), \dots, (b_{n-1},a_{n-1}))=\phi(\g(a_0), \dots,\g(a_{n-1})).
$$
Now we construct a quantum algorithm $A$ with $n$ measurements. We let 
$$
A=((A_l)_{l=0}^{n-1},(b_l)_{l=0}^{n-1},\bar{\varphi}),
$$ 
with $b_0=0$, $b_l\equiv 0\;(1\le l\le n-1)$, and $\bar{\phi}$ as above. Each $A_l$ is of the form
$$
A_l=(Q_l,(U_{lj})_{j=0,1})
$$
with $U_0=U_1=I_{H_m}$ the identity matrix,
$$
Q_l=(m,m',m'', Z_l,\tau,\b),
$$
with $m,m',m'',\b$ as defined above, $Z_l=\{0\}$ and $\tau_l(0)=\l_l$ $(l=0,\dots,n-1)$. This simply means that $A$ is an algorithm which
queries the function $f$  in the appropriate $n$ points, with the needed precision, and  measures the result after each query. Finally
$\bar{\phi}$ is applied.
Let $f\in F$ and let $\zeta$ have distribution $A(f)$. Since the computation remains on the classical states, 
the measurements give  the result
$$
(0,\b(f(\l_0))),\dots,(0,\b(f(\l_{n-1})))
$$
with probability 1.
Hence 
\begin{eqnarray*}
\zeta&=&\bar{\phi}(\b\circ f(\l_0),\dots,\b\circ f(\l_{n-1}))\\
&=&\phi(\g\circ\b\circ f(\l_0),\dots,\g\circ\b\circ f(\l_{n-1}))\\
&=&\phi(\theta\circ f(\l_0),\dots,\theta\circ f(\l_{n-1}))=\wt{A}(\Theta(f))
\end{eqnarray*}
with probability 1, consequently, by conclusion (\ref{RU2}) of Lemma \ref{lem:2} and by (\ref{RU5})
\begin{eqnarray*}
\|S(f)-\zeta\|&\le& \|S(f)-S(\Theta(f))\|+\|S(\Theta(f))-\wt{A}(\Theta(f))\|\\
&\le& \|S\|\delta +2e_n^\de(S,F)+\delta,
\end{eqnarray*}
and (\ref{VU1}) follows.
\end{proof}
From Proposition \ref{pro:4} and Lemma \ref{lem:3} we immediately conclude
\begin{proposition}\label{pro:5}
Let $X$ and $Y$ be normed linear spaces such that $X$ is a linear subspace of $Y$ and the embedding $J:X\to Y$ is continuous. Assume that 
$K=\K$, $F=B_X$, $\Lambda$ is a linearly independent subset of $Y^*$, $T:Y\to G$ is a bounded linear operator, and $S=TJ$. 
Then for all $n_1, n_2\in \N_0$, 
\begin{equation}
\label{CJ4}
e^\q_{\nu_1 n_1+2\nu_2 n_2}(S,F)
\le 
8 e^\de_{n_1}(J,F)\,e^\q_{n_2}(T,B_Y),
\end{equation}
where $\nu_1,\nu_2$ are any numbers satisfying (\ref{KB1}).
If, furthermore, $J$ is a compact operator and $\Lambda$ is linearly independent over $X$, then
\begin{equation}
\label{RT3}
e^\q_{\nu_1 n_1+2\nu_2 n_2}(S,F)
\le 
8e^\de_{n_1}(J,F)\sup_{\mathcal{E}\in\mathscr{C}(B_Y)} e^\q_{n_2}(T, \mathcal{E}).
\end{equation}
\end{proposition}
\section{Weighted mean computation and integration}
Let $L_1^N$, respectively, $L_\infty^N$, be the space of all functions
$f:\Z[0,N)\to \K$, equipped with the norm
$$
\|f\|_{L_1^N}=\frac{1}{N}\sum_{i=0}^{N-1}|f(i)|,
$$
 respectively,
$$
\|f\|_{L_\infty^N}=\max_{0\le i< N}|f(i)|.
$$
We use the notation $L_1^N(\K)$ and $L_\infty^N(\K)$ if the underlying field has to be emphasized. 
Let  $g\in L_1^N$. Define the weighted mean operator $S_{N,g}: L_\infty^N\to \K$ by
$$
S_{N,g}f=\frac{1}{N}\sum_{i=0}^{N-1}g(i)f(i) \quad (f\in L_\infty^N). 
$$
We write $S_{N}$ for $S_{N,g}$ with $g\equiv 1$. We consider the weighted summation problem 
$\mathcal{P}=(B_{L_\infty^N},\K,S_{N,g},\K,\Lambda)$ with $\Lambda=\{\delta_i\,:\,0\le i<N \}$ and 
$\delta_i(f)=f(i)$. Throughout this paper we often use the 
same symbol $c,c_1,\dots$ for possibly different
positive constants (also when they appear in a sequence of relations). 
\begin{proposition}
\label{pro:Q1} There is a constant $c>0$ such that for all $n,N\in \N$ and $g\in L_1^N$
$$
e_n^\q(S_{N,g},B_{L_\infty^N})\le cn^{-1}\|g\|_{ L_1^N}.
$$
\end{proposition}
\begin{proof}
First we consider the case $\K=\R$.
If $g=0$, the statement is trivial. We may assume without loss of generality that $g\ge 0$, otherwise
we split $g$ into its positive and negative part and apply Proposition \ref{pro:3}. Moreover,  by scaling the problem appropriately,
we can assume 
\begin{equation}
\label{QC4}
\|g\|_{L_1^N}=1.
\end{equation}
Now we reduce the problem $S_{N,g}$ 
to the known case $S_M$ for some $M$. Define $h,\tilde{g}\in L_1^N$ by
\begin{equation}
\label{QC5}
h(i)=\lfloor ng(i)\rfloor, \quad \tilde{g}(i)=n^{-1}h(i)\quad(i=0,\dots,N-1).
\end{equation}
We have $|g(i)-\tilde{g}(i)|\le n^{-1}$, therefore
$$
\sup_{f\in B_{L_\infty^N}}|S_{N,g}f-S_{N,\tilde{g}}f|\le n^{-1}.
$$
By Lemma 6 of \cite {Hei01},
\begin{equation}
\label{QC6}
e_n^\q(S_{N,g},B_{L_\infty^N})\le e_n^\q(S_{N,\tilde{g}},B_{L_\infty^N})+n^{-1}.
\end{equation}
Now set $m_0=0$ and for $1\le i\le N$
$$
m_i=\sum_{l=0}^{i-1} h(l),
$$
and denote $m_N=M$. The case $M=0$ is trivial, since this implies $\tilde{g}\equiv 0$, 
thus the result follows directly from (\ref{QC6}). Hence we assume $M\ge 1$. 
Observe that (\ref{QC4}) and (\ref{QC5}) imply 
\begin{equation}
\label{QC7}
M\le nN.
\end{equation}
Define 
$$
\eta:\Z[0,M)\to \Z[0,N)
$$
by $\eta(j)=i$, where $i$ is  the unique integer satisfying $m_i\le j< m_{i+1}$. Let the reduction mapping
$R:L_\infty^N\to L_\infty^M$ be given by
$$
(R(f))(j)=f(\eta(j))\quad (j=0,\dots,M-1).
$$
Clearly, $R$ is of the form (\ref{XG3}), with $\kappa=1$, therefore, by Proposition \ref{pro:1}  we have 
\begin{equation}
\label{KC5}
e_{2n}^\q(S_MR,B_{L_\infty^N})\le e_n^\q(S_M,B_{L_\infty^M}).
\end{equation}
Moreover
\begin{eqnarray*}
S_MR(f)&=&\frac{1}{M}\sum_{j=0}^{M-1}f(\eta(j))=\frac{1}{M}\sum_{i=0}^{N-1}h(i)f(i)\\
&=&\frac{n}{M}\sum_{i=0}^{N-1}\tilde{g}(i)f(i)=\frac{nN}{M}S_{N,\tilde{g}}f.
\end{eqnarray*}
This together with (\ref{QC7}) and (\ref{KC5}) implies 
\begin{eqnarray*}
e_{2n}^\q(S_{N,\tilde{g}},B_{L_\infty^N})&=&e_{2n}^\q\left(\frac{M}{nN}S_M R,B_{L_\infty^N}\right)
=\frac{M}{nN} e_{2n}^\q(S_M R,B_{L_\infty^N})\\
&\le& e_n^\q(S_M,B_{L_\infty^M})
\le cn^{-1},
\end{eqnarray*}
the latter relation being a consequence of \cite{BHM:00} (see also \cite{Hei01}, Theorem 1, for the form stated here).
Combining this with (\ref{QC6}) and scaling the index gives the desired result.

Now we formally derive the complex case from the real case. Let $g\in L_1^N(\C)$ and let $g_1,g_2\in L_1^N(\R)$
be defined by 
\begin{equation}
\label{QV1}
g(j)=g_1(j)+\imath g_2(j)\quad(j=0,\dots,N-1,\;\imath=\sqrt{-1}).
\end{equation}
Clearly, 
\begin{equation}
\label{QV2}
\|g_\a\|_{L_1^N(\R)}\le \|g\|_{ L_1^N(\C)}\quad (\a=1,2).
\end{equation}
We shall express 
$\mathcal{P}=(B_{L_\infty^N},\C,S_{N,g},\K,\Lambda)$ by the help of 
$$
\mathcal{P}=(B_{L_\infty^N},\R,S_{N,g_{\a}},\K,\Lambda)\quad(\a=1,2).
$$
Define $R_1,R_2:L_\infty^N(\C)\to L_\infty^N(\R)$ for $f\in L_\infty^N(\C)$ by
\begin{equation}
\label{QV6}
(R_1f)(j)={\rm Re}(f(j)),\quad (R_2f)(j)={\rm Im}(f(j))\quad(j=0,\dots,N-1).
\end{equation}
Clearly, $R_1,R_2$ are of the form (\ref{XG3}) and map $B_{L_\infty^N(\C)}$ to $B_{L_\infty^N(\R)}$.
Define $J_{\a\b}:\R\to \C$ ($\a,\b\in \{1,2\}$) by
$$
J_{11}a=-J_{22}a=a,\quad J_{12}a=J_{21}a=\imath a \quad (a\in \R).
$$
Then we have, by (\ref{QV1}) and (\ref{QV6}),
\begin{equation*}
S_{N,g}f=\sum_{\a,\b=1}^2 J_{\a\b}S_{N,g_\a}R_\b f.
\end{equation*}
Let $\nu$ be the smallest natural number with $e^{-\nu/8}\le 1/16$. By Proposition \ref{pro:3}
\begin{equation}
\label{QV4}
e_{2\nu n}(S_{N,g},B_{L_\infty^N(\C)})\le 2 \sum_{\a,\b=1}^2 e_{2n}(J_{\a\b}S_{N,g_\a}R_\b,B_{L_\infty^N(\C)}).
\end{equation}
Moreover, by Proposition \ref{pro:1}, 
\begin{equation}
\label{QV5}
e_{2n}(J_{\a\b}S_{N,g_\a}R_\b,B_{L_\infty^N(\C)})\le e_{n}(S_{N,g_\a},B_{L_\infty^N(\R)})\quad (\a,\b=1,2).
\end{equation}
Using the result for the real case and (\ref{QV2}), (\ref{QV4}), and (\ref{QV5}), we  get
$$
e_{2\nu n}(S_{N,g},B_{L_\infty^N(\C)})\le cn^{-1}\|g\|_{L_1^N(\C)},
$$
and a scaling of the index concludes the proof.

\end{proof}
Now we pass to the case of weighted integration. Let $Q\subseteq \R^d$ be a closed, bounded set of positive
Lebesgue measure.  
$L_1(Q)$ denotes the space 
of Lebesgue integrable functions on $Q$ with values in $\K$, equipped 
with the norm
$$
\|f\|_{L_1(Q)}=\int_Q |f(x)|\,dx,
$$
and $L_\infty(Q)$ the space of all $\K$-valued measurable and essentially bounded with respect to the  
Lebesgue measure functions on $Q$, endowed 
with the norm
$$
\|f\|_{\infty}=\mbox{ess\,sup}_{x\in Q} |f(x)|.
$$
Let $g\in L_1(Q)$. Define $I_{Q,g}:L_\infty(Q)\to \K$, the integration operator with weight $g$,  by  
$$
I_{Q,g}f=\int_Q g(x)f(x)\,dx.
$$
$C(Q)$ denotes the space of continuous functions on $Q$, 
equipped with the supremum norm.
A set $\mathcal{E}$ of continuous functions on $Q$
is called uniformly equicontinuous, if  for each
$\e>0$ there is a $\delta>0$ such that for $x,y\in Q$, $|x-y|\le\delta$ implies 
$|f(x)-f(y)|\le \e$ for all $f\in F$. By the Arzel\`{a}-Ascoli theorem, 
bounded, uniformly equicontinuous sets coincide with precompact subsets of $C(Q)$. 
We consider the problem $\mathcal{P}=(B_{C(Q)},\K, I_{Q,g},\K, \Lambda)$ with $\Lambda=\{\delta_x:\, x\in Q\}$, where 
$\delta_x(f)=f(x)$ for $f\in C(Q)$. 
\begin{proposition}
\label{pro:Q2}
There is a constant $c>0$ such
 that for each closed, bounded set $Q\subset \R^d$ of positive Lebesgue measure, for all $g\in L_1(Q)$  and $n\in \N$
$$
\sup_{\mathcal{E}\in\mathscr{C}( B_{C(Q)})}
e_n^\q(I_{Q,g},\mathcal{E})\le cn^{-1}\|g\|_{L_1(Q)}.
$$
\end{proposition}
\noindent {\bf Remark.}  
It is well-known and easily checked by using importance sampling with density function $|g|/\|g\|_{L_1(Q)}$
that in the classical randomized setting we have 
$$
e_n^\ran(I_{Q,g},B_{C(Q)})\le cn^{-1/2}\|g\|_{L_1(Q)},
$$
where $e_n^\ran$ is the $n$-th minimal classical randomized error 
(see, e.g.,  \cite{Hei05b}, section 3). 
Proposition \ref{pro:Q2} is the quantum analogue of this result. 
Let us comment on the reasons for taking the supremum over $\mathcal{E}\in\mathscr{C}( B_{C(Q)})$.
In contrast to the classical randomized setting, 
no non-trivial convergence rate holds for  $e_n^\q(I_{Q,g},B_{C(Q)})$, in general. This is easily checked based
on the fact that a quantum query involves, by definition, the values of functions from $B_{C(Q)}$ 
in a finite set of points of $Q$ only. For situations like this a natural way of formulating quantum counterparts of results 
of the classical randomized setting was already 
observed in section 5 of \cite{Hei01}: If we restrict our analysis to uniformly equicontinuous 
subsets $\mathcal{E}$ of the respective unit ball, non-trivial decay rates can 
be shown in such a way 
that neither the exponent nor the constants involved in these estimates depend on $\mathcal{E}$ (though the number of qubits 
in the respective quantum algorithms does, but 
this is irrelevant for $e_n^\q(I_{Q,g},\mathcal{E})$). 
\medskip

\noindent {\it Proof of Proposition \ref{pro:Q2}.}
Fix  $Q\subset \R^d$, $\mathcal{E}\in\mathscr{C}( B_{C(Q)})$, and $n\in \N$. Let $Q^*$ be a cube with $Q\subseteq Q^*$. 
For $k\in\N$ let 
\begin{eqnarray*}
Q^*= \bigcup_{i=0}^{2^{dk}-1} Q_i^*
\end{eqnarray*}
be the partition of $Q^*$ into $2^{dk}$ congruent cubes of disjoint interior.
Let $Q_i=Q_i^*\cap Q$. Without loss of generality we assume them ordered in such a way that $\mu(Q_i)>0$ iff $i<N$, where
$\mu$ is the Lebesgue measure and $N$ is an appropriate number $1\le N\le 2^{dk}$. Then 
$$
\bigcup_{i=0}^{N-1} Q_i\subseteq Q \quad\mbox{and}\quad \mu\left(Q\setminus \bigcup_{i=0}^{N-1} Q_i\right)=0.
$$
Let $x_i$ be any point in $Q_i$ and
let $P_k$ be the operator of piecewise constant interpolation with respect to the partition
$(Q_i)_{i=0}^{N-1}$ in the points  
$(x_i)_{i=0}^{N-1}$.
By the uniform equicontinuity of $\mathcal{E}$, there is a $k$ such that  
\begin{equation}
\label{K1}
\|f-P_kf\|_{L_\infty(Q)}\le n^{-1}
\end{equation}
for all $f\in \mathcal{E}$. It follows that
\begin{equation}
\label{QG4}
\sup_{f\in\mathcal{E}} |I_{Q,g}f-I_{Q,g}(P_kf)|\le n^{-1}\|g\|_{L_1(Q)}.
\end{equation}
We define 
$R:B_{C(Q)}\to L_\infty^N$ by 
$$
(R(f))(i)=f(x_i)\quad (i=0,\dots,N-1).
$$
Then $R$ is of the form (\ref{XG3}) with $\wt{\Lambda}=\{\delta_i:\, 0\le i<N\}$ and maps $B_{C(Q)}$ to
$B_{L_\infty^N}$. Furthermore, define
$h\in L_1^N$ by $h(i)=N\int_{Q_i}g(y)dy$. Then
\begin{equation*}
\|h\|_{L_1^N}\le \|g\|_{L_1(Q)}.
\end{equation*}
and
\begin{eqnarray}
I_{Q,g}(P_kf)&=& \sum_{i=0}^{N-1}f(x_i)\int_{Q_i}g(t)dt\nonumber\\
  &=&\frac{1}{N}\sum_{i=0}^{N-1}h(i)(R(f))(i)=S_{N,h}\circ R(f).\label{QG5}
\end{eqnarray}
Lemma 6 of \cite{Hei01} together with relations 
(\ref{QG4}) and (\ref{QG5}) imply
\begin{equation}
\label{QE1}
e_n^\q(I_{Q,g},\mathcal{E})\le n^{-1}\|g\|_{L_1(Q)}+e_n^\q(S_{N,h}\circ R,\mathcal{E}).
\end{equation}
By  Propositions \ref{pro:1} and \ref{pro:Q1}, 
$$
e_{2n}^\q(S_{N,h}\circ R,\mathcal{E})
\le e_n^\q(S_{N,h},B_{L_\infty^N})
\le c n^{-1}\|h\|_{L_1^N}\le cn^{-1}\|g\|_{L_1(Q)},
$$
which together with (\ref{QE1}) accomplishes the proof.
\qed

\section{Quantum approximation of weakly singular integral operators}

Let $1\le d_1\le d$ and let $Q_1$ be the closure of an open bounded set in $\R^{d_1}$. We identify $Q_1$ with a subset
of $\R^d$ by identifying $\R^{d_1}$ with $\R^{d_1}\times \{0^{(d-d_1)}\}$. Let $Q_2$ be a bounded Lebesgue measurable
subset of $\R^d$ of positive Lebesgue measure and define ${\rm diag}(Q_1, Q_2):= \{(x,x)\,:\, x\in Q_1\cap Q_2\}$.
We introduce the following class of kernels (see also \cite{Hei05a}, where  
integral operators with such kernels are analyzed).

For $s\in \N$ and
$\s\in \R$ with $-d<\s<+\infty$ we denote by $\mathcal{C}^{s,\s}(Q_1,Q_2)$  the set of all Lebesgue
measurable functions 
$k:Q_1\times Q_2\setminus {\rm diag}(Q_1, Q_2)\to \K$
with the following properties: There is a constant $c> 0$ such that for all $y\in Q_2$
\begin{enumerate}
\item $k(x,y)$ is $s$-times continuously differentiable with respect to $x$ on \newline $Q_1^0\setminus \{y\}$,
where $Q_1^0$ denotes the interior of $Q_1$, considered as a subset of $\R^{d_1}$, 
\item  for all multiindices 
$\a\in N_0^{d_1}$ with $0\le |\a|=\a_1+\dots +\a_{d_1}\le s$ the $\a$-th partial derivative 
$D_x^\a k(x,y)$ of $k$ with respect to the $x$-variables satisfies 
the estimate
\begin{equation}
\label{XF2}
|D_x^\a k(x,y)|\le c\, \left\{\begin{array}{lll}
  |x-y|^{\s-|\a|}+1 & \mbox{if} \quad \s-|\a|\ne 0   \\
  |\ln |x-y||+1 & \mbox{if} \quad \s-|\a|=0
\end{array}
\right. 
\end{equation}
for all $x\in Q_1^0\setminus \{y\}$, and
\item for all $\a\in N_0^{d_1}$ with $0\le |\a|\le s$ the functions $D_x^\a k(x,y)$ 
have continuous extensions to 
$Q_1\setminus\{y\}$. 
\end{enumerate}
We want to extend the definition  to the case $d_1=0$. 
Here we let 
$Q_1=\{0\}\subset \R^d$ and define $\mathcal{C}^{s,\s}(Q_1,Q_2)$  to be the set of all functions 
$k(0,y)$ which are Lebesgue measurable in $y$  
and satisfy 
\begin{equation}
\label{XW2}
|k(0,y)|\le c\, \left\{\begin{array}{lll}
  |y|^{\s}+1 & \mbox{if} \quad \s\ne 0   \\
  |\ln |y||+1 & \mbox{if} \quad \s=0
\end{array}
\right. \quad (y\in Q_2\setminus\{0\})
\end{equation}
with a certain $c>0$. Note that for $d_1=0$ the space $\mathcal{C}^{s,\s}(Q_1,Q_2)$ does not depend 
on $s$. 
 
For $k\in \mathcal{C}^{s,\s}(Q_1,Q_2)$ let $\|k\|_{\mathcal{C}^{s,\s}}$ be the smallest $c>0$ satisfying (\ref{XF2})
or (\ref{XW2}), respectively.
It is easily checked that $\|\, .\,\|_{\mathcal{C}^{s,\s}}$ is a norm on $\mathcal{C}^{s,\s}(Q_1,Q_2)$. 
For $k\in \mathcal{C}^{s,\s}(Q_1,Q_2)$ we
let $T_k$ be the integral operator 
$$
(T_kf)(x)=\int_{Q_2} k(x,y)f(y)dy\quad (x\in Q_1)
$$
acting from $C(Q_2)$ to $L_\infty(Q_1)$ (to $\K$, if $d_1=0$). 
We shall also consider $T_k$ as acting in various other function spaces, 
which will then be mentioned explicitly. 
It is easily checked that $T_k$ maps $C(Q_2)$ into $C(Q_1)$. 
Finally, denote
$$
\mathcal{C}^{\infty,\s}(Q_1,Q_2):=\bigcap_{s\in \N}\mathcal{C}^{s,\s}(Q_1,Q_2).
$$
We start the analysis with the case of $Q_1=[0,1]^{d_1}$, where $0\le d_1\le d$,  and $Q_2$ being a closed
 subset of $[0,1]^d$ of positive Lebesgue measure.
We study the minimal quantum error of approximating  $T_kf$ with $k\in\mathcal{C}^{s,\s}(Q_1,Q_2)$ a fixed kernel,
thus we consider $F=B_{C(Q_2)}$, $G=L_\infty(Q_1)$, $S=T_k$, $K=\K$, and  
$\Lambda=\{\delta_x\,:\,x\in Q_2\}$
(for $d_1=0$ the space $L_\infty(Q_1)$ is replaced by $\K$).

To state the following proposition,
 define $\beta(\s)$ (this parameter will describe the power of the logarithmic term) 
as
\begin{eqnarray}\label{XH3}
\beta(\s) =
\left\{\begin{array}{rll}
 0  & \mbox{if} \quad  \min(s,d+\s,d)>d_1 &  \\[.1cm]
 4  & \mbox{if} \quad \min(s,d+\s,d)=d_1 &  \\  [.1cm]
\frac{\min(s,d+\s)}{d_1}  & \mbox{if} \quad \min(s,d+\s)<d_1 & \quad \mbox{and}\quad s\ne d+\s   \\ [.1cm]
\frac{\min(s,d+\s)}{d_1}+1  & \mbox{if} \quad \min(s,d+\s)<d_1 & \quad \mbox{and}\quad s= d+\s.    \\
    \end{array}
\right.
\end{eqnarray}
Note that, since $d_1\le d$, we have $\min(s,d+\s)<d_1$ iff $\min(s,d+\s,d)<d_1$, so (\ref{XH3}) covers all possible cases.
The following is the quantum version of Proposition 1 of \cite{Hei05a}. For the appearance of the supremum over 
$\mathcal{E}\in\mathscr{C}( B_{C(Q_2)})$ we refer to the remark after Proposition \ref{pro:Q2}.
In the case $d_1=0$ we interpret  $\frac{s}{d_1}=\frac{d+\s}{d_1}=+\infty$.
\begin{proposition}\label{pro:2} Let $0\le d_1\le d$, $s\in \N$, $\s\in \R$, 
$-d<\s<+\infty$, $Q_1=[0,1]^{d_1}$.
Then there is  a constant $c>0$ such that for any closed
subset $Q_2\subseteq [0,1]^d$ of positive Lebesgue measure, and for all $k\in \mathcal{C}^{s,\s}(Q_1,Q_2)$ 
and $n\in \N$ with $n\ge 2$,
\begin{equation}
\label{QS4} 
\sup_{\mathcal{E}\in\mathscr{C}( B_{C(Q_2)})} e_n^\q(T_k,\mathcal{E})
\le c n^{-\min\left(\frac{s}{d_1},\frac{d+\s}{d_1},\,1\right)}(\log n)^{\beta(\s)}\|k\|_{\mathcal{C}^{s,\s}(Q_1,Q_2)},
\end{equation}
where ${\beta(\s)}$ is as defined in (\ref{XH3}).
\end{proposition}

\begin{proof}
In view of Lemma 6 (ii) of \cite{Hei01} it suffices to prove the statement for $k$ with 
\begin{equation}
\label{QT6}
\|k\|_{\mathcal{C}^{s,\s}(Q_1,Q_2)}=1.
\end{equation}
In the case $d_1=0$ we have 
$$
T_kf=\int_{Q_2}k(0,y)f(y)dy,
$$
where by (\ref{QT6}) and (\ref{XW2})
$$
\int_{Q_2}|k(0,y)|dy\le c
$$
(the constants in this proof depend only on $d,d_1,s,\s$), and the result follows directly form Proposition \ref{pro:Q2}. 

Now we assume $d_1\ge 1$. First we recall some notation from \cite{Hei05a}. For $l=0,1,\dots$ let 
\begin{equation}
\label{AX1}
Q_1=\bigcup_{i=1}^{n_l} Q_{1,li}
\end{equation}
be the partition of $Q_1$ into  $n_l=2^{d_1l}$
 closed subcubes of sidelength $2^{-l}$ and mutually disjoint interior.
Let $\Gamma_l$ be the equidistant mesh on $Q_1$ with mesh-size $2^{-l}(\max(s-1,1))^{-1}$, 
$\Gamma_{li}=\Gamma_l\cap Q_{1,li}$ and $\hat{\Gamma}_{li}=\Gamma_{l+1}\cap Q_{1,li}$. 
Let $E_{li}$ be the subspace of $C(Q_{1,li})$ consisting of all
multivariate polynomials on $Q_{1,li}$ of degree at most $\max(s-1,1)$ in each variable. Let $E_l$ be the respective space of 
continuous piecewise polynomial functions on $Q_1$, that is
$$
E_l=\{f\in C(Q_1)\,:\, f|_{Q_{1,li}}\in E_{li},\;i=1,\dots,n_l\}.
$$
Furthermore, define $\hat{E}_{li}\subset C(Q_{1,li})$ by
$$
\hat{E}_{li}=\{f\in C(Q_{1,li})\,:\, f|_{Q_{1,l+1,j}}\in E_{l+1,j}\text{ for all $j$ with } Q_{1,l+1,j}\subset Q_{1,li} \},
$$
in other words, $\hat{E}_{li}\subset C(Q_{1,li})$ is the space of continuous piecewise polynomial functions with respect to the partition of 
$Q_{1,li}$ into subcubes of sidelength $2^{-(l+1)}$.
Let $P_{li}:l_\infty(\Gamma_{li})\to E_{li}$ be the multivariate (tensor product) Lagrange interpolation of degree $\max(s-1,1)$
 on $\Gamma_{li}$,
define $\hat{P}_{li}: l_\infty(\hat{\Gamma}_{li})\to \hat{E}_{li}$ by
$$
(\hat{P}_{li} u)|_{Q_{1,l+1,j}} =  P_{l+1,j} (u|_{\Gamma_{l+1,j}})
$$
for all $j$ with  $Q_{1,l+1,j}\subset Q_{1,li}$, and  $P_l:l_\infty(\Gamma_l)\to E_l$ by
$$
P_lu |_{Q_{1,li}} = P_{li}(u|_{\Gamma_{li}}) \quad (i=1,\dots,n_l)
$$
(continuity follows from the assumption that the degree is $\ge 1$). 
Thus, $\hat{P}_{li}$ and $P_l$ are piecewise Lagrange interpolation operators.
For $f\in C(Q_{1,li})$ or $f\in C(Q_1)$
we write $P_{li}f$ instead of  $P_{li}(f|_{\Gamma_{li}})$, and similarly $\hat{P}_{li}f$ 
and $P_lf$. We shall use the following well-known (see, e.g., \cite{Cia78}) properties: 
 For all $l\in \N_0$ and $i=1,\dots,n_l$, 
\begin{equation}
\label{RA1}
\|P_{li}:l_\infty(\Gamma_{li})\to C(Q_{1,li})\|\le c,
\end{equation}
furthermore, for $f\in C^s(Q_{1,li})$, 
\begin{equation}
\label{RA2}
\|f-P_{li}f\|_{C(Q_{1,li})} \le c\,2^{-sl}\|f\|_{C^s(Q_{1,li})},
\end{equation}
and consequently,
\begin{equation}
\label{RA3}
\|(\hat{P}_{li}-P_{li})f\|_{C(Q_{1,li})} \le c\,2^{-sl}\|f\|_{C^s(Q_{1,li})}.
\end{equation}
Define the embedding operators $J_{li}:C(Q_{1,li})\to L_\infty(Q_1)$ by setting for $x\in Q_1$,
\[
(J_{li}f)(x) \left\{\begin{array}{lll}
  f(x) & \mbox{if} \quad    \\
  0   &\mbox{otherwise.}    \\
    \end{array}
\right. 
\]
We identify $C(Q_1)$ with a subspace of $L_\infty(Q_1)$, thus, the operators $P_l$ can also be considered as acting
into $L_\infty(Q_1)$. First we approximate $T_kf$ by $P_m T_kf$,
where $m\ge 1$ will be fixed later, then $P_m T_kf$ will be approximated by a quantum algorithm. It is readily checked that
\begin{eqnarray}
\label{QA1}
P_m T_kf
&=&P_0T_kf+\sum_{l=0}^{m-1}\sum_{i=1}^{n_l}J_{li}(\hat{P}_{li}-P_{li})T_kf.
\end{eqnarray}
For $l=0,1,\dots m-1$ and $i=1,\dots, n_l$ let $x_{li}$ be the center
and $\rho_l=\sqrt{d_1}2^{-l-1}$ the radius of $Q_{1,li}$.
For $\rho>0$ let $B(x,\rho)$ 
denote the closed $d$-dimensional ball of radius $\rho$ around $x\in \R^{d}$. We represent
\begin{eqnarray}
\label{QA3}
(T_kf)(x)
&=&\int_{B(x_{li},2\rho_l)\cap {Q_2}} k(x,y)f(y)\,dy\nonumber\\
&&+\int_{{Q_2}\setminus B(x_{li},2\rho_l)} k(x,y)f(y)\,dy,
\end{eqnarray}
and introduce $k_{li}\in C(Q_{1,li}\times ({Q_2}\setminus B(x_{li},2\rho_l)))$ by setting for 
$y\in {Q_2}\setminus B(x_{li},2\rho_l)$
\begin{equation}
\label{QS1}
k_{li}(\,\cdot\,,y)=(\hat{P}_{li}-P_{li}) k(\,\cdot\,,y).
\end{equation}
Using that $(\hat{P}_{li}-P_{li})=(\hat{P}_{li}-P_{li})^2$, we conclude
\begin{eqnarray}
\label{KC6}
J_{li}(\hat{P}_{li}-P_{li})T_kf
&=&J_{li}(\hat{P}_{li}-P_{li})\int_{B(x_{li},2\rho_l)\cap {Q_2}} k(\,\cdot\,,y)f(y)\,dy\nonumber\\
&&+J_{li}(\hat{P}_{li}-P_{li})\int_{Q_2\setminus B(x_{li},2\rho_l)} k_{li}(\,\cdot\,,y)f(y)\,dy.
\end{eqnarray}
Next we introduce the following functions: For $x\in \Gamma_0$ define $k_x\in L_1(Q_2)$ as 
$$
k_x(y)=k(x,y) \quad(y\in Q_2),
$$
and for $l=0,\dots,m-1$, $i=1,\dots,n_l$, $x\in \hat{\Gamma}_{li}$ define $g_{lix}, h_{lix}\in L_1(Q_2)$ for
$y\in Q_2$ by
\begin{eqnarray*}
g_{lix}(y)&=&\left\{\begin{array}{lll}
  k(x,y) & \mbox{if}  \quad y \in B(x_{li},2\rho_l)\cap {Q_2}    \\
  0 & \mbox{otherwise,}    \\
    \end{array}
\right. \\[.2cm]
h_{lix}(y)&=&\left\{\begin{array}{lll}
k_{li}(x,y)   & \mbox{if} \quad y \in Q_2\setminus B(x_{li},2\rho_l)    \\
0   & \mbox{otherwise.}    \\
    \end{array}
\right. 
\end{eqnarray*}
Then 
\begin{eqnarray}
\label{KC7}
P_mT_kf&=&P_0\left((I_{Q_2,k_x}f)_{x\in \Gamma_0}\right)\nonumber\\
&&+\sum_{l=0}^{m-1}\sum_{i=1}^{n_l}J_{li}(\hat{P}_{li}-P_{li})
\left(\left(I_{Q_2,g_{lix}}f+I_{Q_2,h_{lix}}f\right)_{x\in\hat{\Gamma}_{li}}\right).
\end{eqnarray}
From (\ref{QT6}) and (\ref{XF2}) we have
\begin{equation}
\label{QA2}
\|k_x\|_{L_1(Q_2)}=\int_{Q_2} |k(x,y)|\,dy\le c  \quad (x\in \Gamma_0).
\end{equation}
For $x\in \hat{\Gamma}_{li}$ we deduce from (\ref{XF2}) that 
\begin{eqnarray}
\|g_{lix}\|_{L_1(Q_2)}&=&\int_{B(x_{li},2\rho_l)\cap {Q_2}} |k(x,y)|\,dy\nonumber\\
&\le& \int_{B(x,3\rho_l)\cap {Q_2}}|k(x,y)|\,dy
\nonumber\\
&\le &c\int_{B(x,3\rho_l)}\left(|x-y|^\s+|\ln |x-y|| +1\right) dy\nonumber\\
&\le & c\,(2^{-(d+\sigma)l}+(l+1)2^{-dl}).
\label{QA4}
\end{eqnarray}
Furthermore, again by (\ref{XF2}), we have for $x\in Q_{1,li}$ and $y\in Q_2\setminus B(x_{li},2\rho_l)$ 
\begin{eqnarray}
\nonumber\\
|D^\a_x k(x,y)|&\le& c\,(|x-y|^{\s-|\a|}+|\ln |x-y|| +1)\\
&\le& c\,(|x_{li}-y|^{\s-|\a|}+|\ln |x_{li}-y|| +1),\nonumber
\end{eqnarray}
hence 
\begin{equation}
\label{QA5}
\|k(\,.\,,y)\|_{C^s(Q_{1,li})}\le c\,(|x_{li}-y|^{\s-s}+|\ln |x_{li}-y|| +1).
\end{equation}
Using (\ref{RA3}) and (\ref{QS1}) we obtain for $y\in Q_2\setminus B(x_{li},2\rho_l)$ 
\begin{equation}
\label{QS2}
\|k_{li}(\,.\,,y)\|_{C(Q_{1,li})}\le c\,2^{-sl}(|x_{li}-y|^{\s-s}+|\ln |x_{li}-y|| +1).
\end{equation}
We have 
$$
\int_{Q_2\setminus B(x_{li},2\rho_l)}|x_{li}-y|^{\s-s}dy\le c
\left\{\begin{array}{lll}
 1  & \mbox{if} \quad  \s-s>-d  \\
 l+1  & \mbox{if} \quad  \s-s=-d  \\
 2^{-(\s-s+d)l}  & \mbox{if} \quad  \s-s<-d. \\
    \end{array}
\right. 
$$
Therefore, integrating (\ref{QS2}), we get for $x\in\hat{\Gamma}_{li}$
\begin{equation}
\label{QS3}
\|h_{lix}\|_{L_1(Q_2)}=\int_{Q_2\setminus B(x_{li},2\rho_l)}|k_{li}(x,y)|dy\le c\,(l+1)^{\a_0} 2^{-\min(s,d+\s)l},
\end{equation}
where
\begin{equation}
\label{QA6}
\a_0=\left\{\begin{array}{lll}
 1  & \mbox{if} \quad s=d+\s   \\
 0  & \mbox{otherwise.}    \\
    \end{array}
\right. 
\end{equation}
Now we approximate the integrals in (\ref{KC7}) by quantum algorithms. Let $\mathcal{E}\in\mathscr{C}(B_{C(Q_2)})$ (as already mentioned, 
the constants depend only on the parameters $d,d_1,s,\s$, and in particular not on $\mathcal{E}$).
 Using Proposition \ref{pro:Q2} together with (\ref{QA2}), (\ref{QA4}), and (\ref{QS3}), we obtain the following relations
\begin{equation}
\label{QA7}
e_{N^{(0)}}^\q(I_{Q_2,k_x},\mathcal{E})\le c/N^{(0)} \quad (x\in \Gamma_0)
\end{equation}
\begin{equation}
\label{QA8}
e_{N_l}^\q(I_{Q_2,g_{lix}},\mathcal{E})\le c(2^{-(d+\sigma)l}+(l+1)2^{-dl})N_l^{-1} 
\quad (x\in\hat{\Gamma}_{li})
\end{equation}
\begin{equation}
\label{QA9}
e_{N_l}^\q(I_{Q_2,h_{lix}},\mathcal{E})\le c\,(l+1)^{\a_0} 2^{-\min(s,d+\s)l}N_l^{-1}
\quad (x\in\hat{\Gamma}_{li}),
\end{equation}
where $N^{(0)}, N_l\in \N\; (l=0,\dots,m-1)$ are arbitrary natural numbers which will be fixed later. 
Let $\nu^{(0)},\nu_l\in \N\;(l=0,\dots,m-1)$ be the smallest natural numbers satisfying
\begin{eqnarray}
|\Gamma^{(0)}|e^{-\nu^{(0)}/8}&\le& 2^{-3}\label{QB0}\\
2\sum_{i=1}^{n_l}|\hat{\Gamma}_{li}|e^{-\nu_l/8}&\le&  2^{-(l+4)}\quad(l=0,\dots,m-1).\label{QB0A}
\end{eqnarray}
Consequently
\begin{equation}
\label{QC8}
\nu_l\le c(l+1)\quad(l=0,\dots,m-1)
\end{equation}
and 
\begin{equation}
\label{QG1}
|\Gamma^{(0)}|e^{-\nu^{(0)}/8}+2\sum_{l=0}^{m-1}\sum_{i=1}^{n_l}|\hat{\Gamma}_{li}|e^{-\nu_l/8}\le 1/4.
\end{equation}
By Lemma 3 of \cite{Hei01}
we can assert the existence of quantum algorithms $A^{(0)}_{x}\; (x\in \Gamma_0)$
with
\begin{equation}
\label{QB1}
e(I_{Q_2,k_x},A^{(0)}_{x},\mathcal{E},e^{-\nu^{(0)}/8})\le c/N^{(0)},\quad n_q(A^{(0)}_{x})\le \nu^{(0)} N^{(0)} \quad (x\in \Gamma_0)
\end{equation}
and of quantum algorithms $A^{(1)}_{lix}$ and $A^{(2)}_{lix}$  $(x\in\hat{\Gamma}_{li}$, $i=1\dots n_l$, $l=0,\dots,m-1$), 
such that 
\begin{equation}
\label{QB4}
n_q(A^{(1)}_{lix})\le \nu_l N_l,\qquad n_q(A^{(2)}_{lix})\le \nu_l N_l
\end{equation}
\begin{equation}\label{QB2}
e(I_{Q_2,g_{lix}},A^{(1)}_{lix},\mathcal{E},e^{-\nu_l/8}) 
\le c(2^{-(d+\sigma)l}+(l+1)2^{-dl})N_l^{-1}
\end{equation}
and
\begin{equation}
\label{QB3}
e(I_{Q_2,h_{lix}},A^{(2)}_{lix},\mathcal{E},e^{-\nu_l/8}) 
\le c\,(l+1)^{\a_0} 2^{-\min(s,d+\s)l}N_l^{-1}
\end{equation}
for all $x\in\hat{\Gamma}_{li}$. We define the quantum algorithm  
\begin{equation}
\label{QB7}
A=P_0\left(\left(A^{(0)}_{x}\right)_{x\in \Gamma_0}\right)+
\sum_{l=0}^{m-1}\sum_{i=1}^{n_l}J_{li}(\hat{P}_{li}-P_{li})\left(\left(A^{(1)}_{lix}+A^{(2)}_{lix}\right)_{x\in\hat{\Gamma}_{li}}\right)
\end{equation}
in the sense of the composition of quantum algorithms described in \cite{Hei01}, relation (11). Let $f\in \mathcal{E}$
and let $\zeta^{(0)}_{x},\zeta^{(1)}_{lix}, \zeta^{(2)}_{lix}$ ($x\in\hat{\Gamma}_{li}$, $i=1\dots n_l$, $l=0,\dots,m-1$) 
be independent random variables with distribution $A^{(0)}_{x}(f)$, $A^{(1)}_{lix}(f)$, and $A^{(2)}_{lix}(f)$, respectively.
From Lemma 2  of \cite{Hei01} it follows that the random variable
\begin{equation}
\label{QC1}
\zeta=P_0\left(\left(\zeta^{(0)}_{x}\right)_{x\in \Gamma_0}\right)+
\sum_{l=0}^{m-1}\sum_{i=1}^{n_l}J_{li}(\hat{P}_{li}-P_{li})
\left(\left(\zeta^{(1)}_{lix}+\zeta^{(2)}_{lix}\right)_{x\in\hat{\Gamma}_{li}}\right)
\end{equation}
has distribution $A(f)$, and that
\begin{equation}
\label{QC9}
n_q(A)=\sum_{x\in \Gamma^{(0)}}   n_q\left(A^{(0)}_{x}\right)+\sum_{l=0}^{m-1}\sum_{i=1}^{n_l}\sum_{x\in\hat{\Gamma}_{li}}
\left( n_q\left(A^{(1)}_{lix}\right)+n_q\left(A^{(2)}_{lix}\right)\right)
\end{equation}
By (\ref{KC7}), (\ref{QC1}),  and (\ref{RA1}),
\begin{eqnarray}
\label{KC7A}
\lefteqn{  \|T_kf - \zeta\| }\nonumber\\
&\le&\left\|T_kf-P_mT_kf\right\| +\left\|P_0\left(\left(I_{Q_2,k_x}f-\zeta^{(0)}_{x}\right)_{x\in \Gamma_0}\right)\right\|\nonumber\\
&+&\sum_{l=0}^{m-1}\sum_{i=1}^{n_l}\left\|J_{li}(\hat{P}_{li}-P_{li})\left(\left(I_{Q_2,g_{lix}}f-\zeta^{(1)}_{lix}
\right)_{x\in\hat{\Gamma}_{li}}\right)\right.\nonumber\\
&&\qquad \qquad+\left. J_{li}(\hat{P}_{li}-P_{li})\left(\left(
I_{Q_2,h_{lix}}f-\zeta^{(2)}_{lix}\right)_{x\in\hat{\Gamma}_{li}}\right)\right\|\nonumber\\
&\le&\|T_kf-P_mT_kf\|+c\max_{x\in \Gamma_0}\left|I_{Q_2,k_x}f-\zeta^{(0)}_{x}\right|\nonumber\\
&+&c\sum_{l=0}^{m-1}\sum_{i=1}^{n_l}\left(\max_{x\in\hat{\Gamma}_{li}} \left|I_{Q_2,g_{lix}}f-\zeta^{(1)}_{lix}\right|
+\max_{x\in\hat{\Gamma}_{li}} \left|I_{Q_2,h_{lix}}f-\zeta^{(2)}_{lix}\right|\right).
\end{eqnarray}
As established in \cite{Hei05a}, Lemma 3, the error of approximation by $P_mT_kf$ satisfies
\begin{equation}
\label{QA10}
\|T_kf-P_mT_kf\|\le c\,(m^{\a_0} 2^{-\min(s,d+\s)m}+m\,2^{-dm}).
\end{equation}
Furthermore, from  (\ref{QG1}), (\ref{QB1}), (\ref{QB2}), and (\ref{QB3}), we conclude that with probability
at least $3/4$ the following relations hold simultaneously:
\begin{equation}
\label{QB8}
\max_{x\in \Gamma_0}|I_{Q_2,k_x}f-\zeta^{(0)}_{x}|\le c/N^{(0)}
\end{equation}
and for $i=1\dots n_l$, $l=0,\dots,m-1$ 
\begin{equation}
\label{QB9}
\max_{x\in\hat{\Gamma}_{li}} |I_{Q_2,g_{lix}}f-\zeta^{(1)}_{lix}|
\le c(2^{-(d+\sigma)l}+(l+1)2^{-dl})N_l^{-1}
\end{equation}
and
\begin{equation}
\label{QB10}
\max_{x\in\hat{\Gamma}_{li}} |I_{Q_2,h_{lix}}f-\zeta^{(2)}_{lix}|\le c\,(l+1)^{\a_0} 2^{-\min(s,d+\s)l}N_l^{-1}.
\end{equation}
We get
from (\ref{KC7A}--\ref{QB10})
\begin{eqnarray}\label{QA11}
\lefteqn{e(T_k,A,\mathcal{E})}\nonumber\\[.2cm]
&\le& c(m^{\a_0}2^{-\min(s,d+\s)m}+m\,2^{-dm})+c\big(N^{(0)}\big)^{-1}+\nonumber\\
&& c\sum_{l=0}^{m-1}N_l^{-1}\left((l+1)^{\a_0}2^{-\min(s,d+\s)l}+(l+1)2^{-dl}\right).
\end{eqnarray}
By (\ref{QC8}), (\ref{QB1}), (\ref{QB4}), and (\ref{QC9}), the number of quantum queries of $A$ satisfies 
\begin{eqnarray}\label{KA1}
n_q(A)&\le& 
\nu^{(0)}N^{(0)}|\Gamma^{(0)}|+2\sum_{l=0}^{m-1}\sum_{i=1}^{n_l}\nu_l N_l|\hat{\Gamma}_{li}|\nonumber\\
&\le& c\left(N^{(0)}+\sum_{l=0}^{m-1} (l+1) 2^{d_1 l}N_l\right).
\end{eqnarray}
Let $n\in \N$ with $n\ge 2$ be given. 
First we consider the case $\min(s,d+\s,d)>d_1$. Let $\tau>0$ be such that 
$$
\min(s,d+\s,d)>d_1+\tau,
$$
 and put 
$$
m=\left\lceil\frac{\log n}{d_1+\tau}\right\rceil
$$
($\log$ always means $\log_2$),
$$
N^{(0)}= n,\quad N_l=\left\lceil n\,2^{-(d_1+\tau)l}\right\rceil\quad (l=0,\dots,m-1).
$$
By (\ref{QA11}) we get for the quantum error:
\begin{eqnarray*}
\lefteqn{e(T_k,A,\mathcal{E})}\\
&\le& c(m^{\a_0}2^{-\min(s,d+\s)m}+m\,2^{-dm}+n^{-1})\\
&&+c\sum_{l=0}^{m-1} n^{-1}2^{(d_1+\tau)l}
\left((l+1)^{\a_0}  2^{-\min(s,d+\s)l}+(l+1)2^{-dl}\right)\\ 
&\le& c n^{-1}.
\end{eqnarray*}
It follows from (\ref{KA1}) that the number of quantum queries satisfies
\begin{eqnarray*}
n_q(A)&\le& c\left(
n+\sum_{l=0}^{m-1}(l+1) 2^{d_1l}( n 2^{-(d_1+\tau)l}+1)\right)\nonumber\\
&\le& c(n+m2^{d_1m}) \le cn,
\end{eqnarray*}
and a simple change of variables in the index yields the desired result (\ref{QS4}) for this case.

Next assume $\min(s,d+\s,d)=d_1$, put 
\begin{equation}
\label{QA12}
m=\left\lceil \frac{\log n}{d_1}\right\rceil,
\end{equation}
and 
$$
N^{(0)}=n,\quad N_l=\left\lceil nm^{-1}2^{-d_1l}\right\rceil\quad (l=0,\dots,m-1).
$$
Then we have, by (\ref{QA11}), 
\begin{eqnarray*}
\lefteqn{e(T_k,A,\mathcal{E})}\\
&\le& c (m^{\a_0} 2^{-d_1 m}+m2^{-dm}+n^{-1})+\\
&&c\sum_{l=0}^{m-1}mn^{-1}2^{d_1l}
\left((l+1)^{\a_0}2^{-\min(s,d+\s)l}+(l+1) 2^{-dl}\right)
\\
&\le& c mn^{-1} \sum_{l=0}^{m-1}(l+1)\\
&\le& cn^{-1}m^3\le cn^{-1}(\log n)^3,
\end{eqnarray*}
and, by (\ref{KA1}),
\begin{eqnarray*}
n_q(A)&\le& c\left(n+\sum_{l=0}^{m-1} (l+1)2^{d_1l}(nm^{-1}2^{-d_1 l}+1)\right)\\ 
&\le& c (mn + m2^{d_1m})\le c\,n\log n.
\end{eqnarray*}
Again we get (\ref{QS4}) by a change of variables in the index.
 
Finally, we suppose $\min(s,d+\s)<d_1$. Let $\tau>0$ be such that 
$$
\min(s,d+\s)+\tau<d_1,
$$ and put 
\begin{equation}
\label{QA13}
m=\left\lceil \frac{\log n}{d_1}\right\rceil,
\end{equation}
$$
N^{(0)}=n,\quad N_l=\left\lceil   n\,2^{-d_1l-\tau(m-l)} \right\rceil \quad (l=0,\dots,m-1).
$$
We derive from (\ref{QA11})
\begin{eqnarray*}
\lefteqn{e(T_k,A,\mathcal{E})}\\[.2cm]
&\le& c(m^{\a_0} 2^{-\min(s,d+\s)m}+m 2^{-dm}+n^{-1})+\\
&&c\sum_{l=0}^{m-1}n^{-1}2^{d_1l+\tau(m-l)}
\Big((l+1)^{\a_0}2^{-\min(s,d+\s)l}+(l+1)2^{-dl}\Big)\\
&\le&cn^{-\min(s,d+\s)/d_1}(\log n)^{\a_0}+\\
&&c\sum_{l=0}^{m-1}n^{-1}2^{d_1l+\tau(m-l)}
(l+1)^{\a_0}2^{-\min(s,d+\s)l}\\
&\le&cn^{-\min(s,d+\s)/d_1}(\log n)^{\a_0}\\
&&+cn^{-1}m^{\a_0}2^{\tau m}\sum_{l=0}^{m-1}2^{(d_1-\tau-\min(s,d+\s))l}\\
&\le&cn^{-\min(s,d+\s)/d_1}(\log n)^{\a_0}+cn^{-1}m^{\a_0} 2^{(d_1-\min(s,d+\s))m}\\
&\le&cn^{-\min(s,d+\s)/d_1}(\log n)^{\a_0}.
\end{eqnarray*}
The number of queries is
\begin{eqnarray*}
n_q(A)&\le& c\left(n+\sum_{l=0}^{m-1}(l+1) 2^{d_1l}(n2^{-d_1 l-\tau(m-l)}+1)\right) \\ 
&\le& c\left(n\sum_{l=0}^{m-1}(l+1)2^{-\tau(m-l)}+m2^{d_1m}\right)\le c\, n\log n,
\end{eqnarray*}
and (\ref{QS4}) follows. This concludes the proof of Proposition \ref{pro:2}.
\end{proof}

Next we consider the case  $Q_1=[0,1]^{d_1}$ and $Q_2=[0,1]^d$, where $0\le d_1\le d$.
Again we study the approximation of $T_kf$ with $k\in\mathcal{C}^{s,\s}(Q_1,Q_2)$ a fixed kernel, but now
$f\in C^r(Q_2)$, and thus, the operator $T_k$ is considered as acting from $C^r(Q_2)$ to $L_\infty(Q_1)$.
Here $r\in \N$ and $C^r(Q_2)$ denotes
the space of continuous complex-valued functions on $Q$ which are $r$-times continuously differentiable in the interior 
$Q_2^0$, and whose partial derivatives up to order $r$ have continuous extensions to $Q_2$. The norm on $C^r(Q_2)$
is defined as
$$
\|f\|_{C^r(Q_2)}=\max_{|\a|\le r} \sup_{x\in Q_2}|D^\a f(x)|.
$$
We let $F=B_{C^r(Q_2)}$, $G=L_\infty(Q_1)$, $S=T_k$, $\K=\K$, and 
\begin{equation}
\label{QH1}
\Lambda=\{\delta_x\,:\,x\in Q_2\},
\end{equation}
where $\delta_x(f)= f(x)$. 

To cover the logarithmic factors, we introduce for $\s\in\R$ with $-d<\s<+\infty$
\begin{equation}
\label{AL1}
\kappa(\s)=\left\{\begin{array}{llr}
  0 & \mbox{if}& \min(d+\s,d)>d_1  \\
  4  & \mbox{if} &\min(d+\s,d)=  d_1   \\
 \frac{d+\s}{d_1} & \mbox{if}& d+\s<d_1=d,
    \end{array}
\right. 
\end{equation}
and if $d+\s<d_1<d$, we fix any $\epsilon_0>0$ and define
\begin{equation}
\label{AL6}
\kappa(\s)=\left\{\begin{array}{lll}
 4 & \mbox{if} \quad \frac{r+d+\s}{d_1}< \frac{r}{d}+1  \\
  \frac{r}{d}+6+\epsilon_0 & \mbox{if} \quad \frac{r+d+\s}{d_1}= \frac{r}{d}+1\\
 4+ \epsilon_0& \mbox{if} \quad  \frac{r+d+\s}{d_1}> \frac{r}{d}+1.  
    \end{array}
\right. 
\end{equation}
\begin{proposition}\label{pro:3a} Assume $0\le d_1\le d$, $s\in \N$, $s> d_1$, $\s\in \R$, 
$-d<\s<+\infty$, $r\in \N$.
Then there is  a constant $c>0$ such that for all $k\in B_{\mathcal{C}^{s,\s}(Q_1,Q_2)}$ and $n\in \N$ with $n\ge 2$,
\begin{equation}
\label{QH2}
e_n^\q(T_k,B_{C^r(Q_2)})
\le c n^{-\min\left(\frac{r+d+\s}{d_1},\,\frac{r}{d}+1\right)}(\log n)^{\kappa(\s)}.
\end{equation}
where ${\kappa(\s)}$ is as defined in (\ref{AL1}), (\ref{AL6}).
\end{proposition}
\begin{proof} Let $n\in \N$, $n\ge 2$. Let $\nu$ be the smallest natural number such that 
\begin{equation}\label{IA2}
e^{-\nu/8}\le 1/8.
\end{equation}
First we assume that either
\begin{equation}
\label{QH3}
d+\s\ge d_1
\end{equation}
(which, because of $d_1\le d$, is equivalent to $\min(d+\s,d)\ge d_1$) or 
\begin{equation}
\label{QH4}
d+\s<d_1=d.
\end{equation}
Comparing (\ref{AL1}) with (\ref{XH3}), we conclude that in these cases
\begin{equation}
\label{QH5}
\kappa(\s)=\beta(\s).
\end{equation}
We write $T_k=\bar{T}_k J$, with
$J$ the identical embedding $C^r(Q_2)\to C(Q_2)$, and
$\bar{T}_k$ the operator $T_k$, considered as acting from $C(Q_2)$ to $L_\infty(Q_1)$. 
With $X=C^r(Q_2)$, $Y=C(Q_2)$, and $\Lambda$ as given by (\ref{QH1}), the assumptions of Proposition \ref{pro:5} are 
easily verified. Therefore
\begin{equation}
\label{QG7}
e_{3\nu n}^\q(T_k,B_{C^r(Q_2)})\le 8e_{n}^\de(J,B_{C^r(Q_2)})\sup_{\mathcal{E}\in\mathscr{C}(B_{C(Q_2)})}e_{n}^\q(\bar{T}_k,\mathcal{E}).
\end{equation}
By Proposition \ref{pro:2} and (\ref{QH5}),
\begin{equation}
\label{QG8}
\sup_{\mathcal{E}\in\mathscr{C}( B_{C(Q_2)})} e_{n}^\q(\bar{T}_k,\mathcal{E})\le cn^{-\min\left(\frac{d+\s}{d_1},1\right)}
(\log n)^{\kappa(\s)}
\end{equation}
(the constants in this proof depend only on $d,d_1,r,s,\s$).
It is well-known that
\begin{equation}
\label{QG9}
e_{n}^\de(J,B_{C^r(Q_2)})\le cn^{-\frac{r}{d}}.
\end{equation}
Furthermore, if (\ref{QH3}) or (\ref{QH4}) holds, we have
$$
\frac{r}{d}+ \min\left(\frac{d+\s}{d_1},1\right) = \min\left(\frac{r+d+\s}{d_1},\,\frac{r}{d}+1\right).
$$
This  together with (\ref{QG7}), (\ref{QG8}), and (\ref{QG9}) implies the desired result.

Now we assume $d+\s<d_1<d$ (hence $d_1\ne 0$). We recall the following construction from \cite{Hei05b}, proof 
of Proposition 4: We decompose $Q_2=\bigcup_{l=0}^{m}H_l$ with
\begin{equation}
\label{AK10}
m=\left\lceil\frac{\log n}{d_1}\right\rceil,
\end{equation}
$$
H_l=[0,1]^{d_1}\times\left(2^{-l}[0,1]^{d-d_1}\setminus 2^{-(l+1)}[0,1)^{d-d_1}\right)\quad(l=0,\dots,m-1)
$$
and
$$
H_m= [0,1]^{d_1}\times\left(2^{-m}[0,1]^{d-d_1}\right).
$$
Let
$$
k_l(x,y)=k(x,y)\quad(x\in Q_1,\, y\in H_l).
$$
Clearly, $k_l\in\mathcal{C}^{s,\s}(Q_1,H_l)$ and
\begin{equation}
\label{B3}
\|k_l\|_{\mathcal{C}^{s,\s}(Q_1,H_l)}\le  \|k\|_{\mathcal{C}^{s,\s}(Q_1,Q_2)}.
\end{equation}
Put $\s_1=d_1-d$. Arguing as in \cite{Hei05b}, proof of relation (52), we conclude
\begin{equation}
\label{AK14}
\|k_l\|_{\mathcal{C}^{s,\s_1}(Q_1,H_l)}\le c 2^{(\s_1-\s)l}\quad (0\le l<m).
\end{equation}
We have the following representation
\begin{equation}
\label{B4}
T_k=\sum _{l=0}^mT_{k_l}J_lR_l,
\end{equation}
where $R_l:C^r(Q_2)\to C^r(H_l)$ is the restriction operator,
$J_l:C^r(H_l)\to C(H_l)$ is the embedding, and $T_{k_l}$ is considered
as an operator from $C(H_l)$ to $L_\infty(Q_1)$.
With real numbers $\delta_1, \delta_2\ge 0$, which will be defined later, we put 
\begin{equation}
\label{AK13}
p_l=\left\lceil 2^{\left(\frac{d_1}{d}-\delta_1\right)(m-l)-\delta_2 l}\right\rceil  \quad (0\le l\le m).
\end{equation}
Observe that $p_m=1$. Furthermore, define
\begin{equation}
\label{BE1}
n_l=2^{d_1 (l+1)}p_l^{d}  \quad (0\le l\le m). 
\end{equation}
Note that  $n_l\ge 2$ for $0\le l\le m$. 
As shown in \cite{Hei05b}, proof of Proposition 4, there is a constant $c_1\in \N$ such that 
\begin{equation}
\label{AK4}
e_{c_1n_l}^\de(J_l, B_{C^r(H_l)})\le c\,2^{-rl}p_l^{-r} \quad (0\le l\le m).
\end{equation}
We verify that for $0\le l\le m$
\begin{equation}
\label{AK5}
\sup_{\mathcal{E}\in\mathscr{C}( B_{C(H_l)})} e_{n_l}^\q(T_{k_l},\mathcal{E})
\le c2^{-(d+\s)l} p_l^{-d}(\log n_l)^4.
\end{equation}
Indeed, in the case $0\le l<m$ relation (\ref{AK14}) and Proposition \ref{pro:2} yield
\begin{eqnarray*}
\sup_{\mathcal{E}\in\mathscr{C}( B_{C(H_l)})} e_{n_l}^\q(T_{k_l},\mathcal{E})&\le& c n_l^{-1} 2^{(\s_1-\s)l} 
(\log n_l)^4\nonumber\\
&\le& c 2^{-d_1l} p_l^{-d} 2^{(d_1-d-\s)l}(\log n_l)^4\nonumber\\
&=&c2^{-(d+\s)l} p_l^{-d}(\log n_l)^4.
\end{eqnarray*}
If $l=m$, (\ref{B3}) and Proposition \ref{pro:2} give
$$
\sup_{\mathcal{E}\in\mathscr{C}( B_{C(H_l)})} e_{n_m}^\q(T_{k_m},\mathcal{E})
\le cn_m^{-\frac{d+\s}{d_1}}(\log n_m)^{\frac{d+\s}{d_1}}\le c2^{-(d+\s)m}(\log n_m)^{\frac{d+\s}{d_1}},
$$
and, since $p_m=1$ and $\frac{d+\s}{d_1}<1$,  (\ref{AK5}) follows.
For $l=0,\dots,m$ we set
\begin{equation}
\label{I1A} 
 \nu_l=\left\lceil 8(2\ln(m-l+1)+ \ln 8)\right\rceil. 
\end{equation}
It follows from (\ref{I1A}) that 
\begin{equation}
\label{I2}
\sum_{l=0}^{m}e^{-\nu_l/8}\le \frac{1}{8}\sum_{l=0}^{m}(m-l+1)^{-2}<\frac{1}{4}.
\end{equation}
Define
$$
\bar{n}=2(c_1+2)\nu\sum_{l=0}^m \nu_ln_l
$$
 with $c_1$ from (\ref{AK4}) and $\nu$ from (\ref{IA2}). Then (\ref{B4}), (\ref{I2}) and Proposition
\ref{pro:3} imply
\begin{eqnarray}
 e_{\bar{n}}^\q(T_k,B_{C^r(Q_2)})
&\le&2\sum_{l=0}^m e_{2(c_1+2)\nu n_l}^\q(T_{k_l}J_lR_l,B_{C^r(Q_2)}).\nonumber
\end{eqnarray}
The mapping $R_l$ is of the form (\ref{XG3}) with $\Lambda=\{\delta_x:\, x\in Q_2\}$, $\wt{\Lambda}=\{\delta_x:\, x\in H_l\}$, 
$\kappa=1$, and satisfies $R_l(B_{C^r(Q_2)})\subseteq B_{C^r(H_l)}$, hence,
by Proposition \ref{pro:1},
$$
e_{2(c_1+2)\nu n_l}^\q(T_{k_l}J_lR_l,B_{C^r(Q_2)})\le e_{(c_1+2)\nu n_l}^\q(T_{k_l}J_l,B_{C^r(H_l)}).
$$
Furthermore, by Proposition \ref{pro:5},
$$
e_{(c_1+2)\nu n_l}^\q(T_{k_l}J_l,B_{C^r(H_l)})\le 8e_{c_1n_l}^\de(J_l,B_{C^r(H_l)})
\sup_{\mathcal{E}\in\mathscr{C}(B_{C(H_l)})}e_{n_l}^\q(T_{k_l},\mathcal{E}).
$$
Using this and (\ref{AK13}--\ref{AK5}), we get
\begin{eqnarray}
\lefteqn{e_{\bar{n}}^\q(T_k,B_{C^r(Q_2)}) }\nonumber\\
&\le&16\sum_{l=0}^m e_{c_1n_l}^\de(J_l,B_{C^r(H_l)})\sup_{\mathcal{E}\in\mathscr{C}(B_{C(H_l)})}e_{n_l}^\q(T_{k_l},\mathcal{E}) \nonumber\\
&\le& c\sum_{l=0}^m 2^{-(r+d+\s)l} p_l^{-\left(r+d\right)}(\log n_l)^4\nonumber\\
&\le& cm^4\sum_{l=0}^m 
2^{-\left(r+d+\s-\delta_2\left(r+d\right)\right)l
-\left(r+d\right) \left(\frac{d_1}{d}-\delta_1\right)(m-l)}.\label{AK9}
\end{eqnarray}
Relations (\ref{IA2}), (\ref{AK10}), (\ref{AK13}), (\ref{BE1}), and (\ref{I1A}), give
\begin{equation*}
\bar{n}=2(c_1+2)\nu\sum_{l=0}^m  \nu_l n_l\le c \sum_{l=0}^m \nu_l 2^{d_1l}
\left( 2^{d_1(m-l)-\delta_1 d (m-l)-\delta_2 dl}+1\right),
\end{equation*}
therefore, if $\delta_1>0$,
\begin{equation}
\label{AK8A}
\bar{n}\le c2^{d_1m}\le cn,
\end{equation}
if $\delta_2>0$,
\begin{equation}
\label{AK8B}
\bar{n}\le c\,2^{d_1m}\log(m+1)\le cn\log\log (n+1),
\end{equation}
and if $\delta_1=\delta_2=0$,
\begin{equation}
\label{AK8C}
\bar{n}\le c \,2^{d_1m}m\log(m+1)\le cn \log n \log\log (n+1).
\end{equation}
The proof will be accomplished by considering three cases. 
The first case is $r+d+\s<(r+d)\frac{d_1}{d}$, that is, $\frac{r+d+\s}{d_1}<\frac{r}{d}+1$. Here we put $\delta_2=0$ and take any 
$\delta_1>0$ satisfying
$$
r+d+\s<\left(r+d\right) \left(\frac{d_1}{d}-\delta_1\right).
$$
From (\ref{AK9}) and  (\ref{AK10}),
$$
e_{\bar{n}}^\q(T_k,B_{C^r(Q_2)})\le c m^4 2^{-\left(r+d+\s\right)m}
\le c n^{-\frac{r+d+\s}{d_1}}(\log n)^4.
$$
Relation (\ref{AK8A}) and a suitable scaling lead to 
$$
e_n^\q(T_k,B_{C^r(Q_2)})\le c n^{-\frac{r+d+\s}{d_1}}(\log n)^4.
$$
The next case is  $r+d+\s=(r+d)\frac{d_1}{d}$. Here we put $\delta_1=\delta_2=0$, and obtain from (\ref{AK9}),
\begin{eqnarray*}
e_{\bar{n}}^\q(T_k,B_{C^r(Q_2)})&\le& cm^5 2^{-\left(r+d\right)\frac{d_1}{d}m}
\le cn^{-\left(\frac{r}{d}+1\right)}(\log n)^5.
\end{eqnarray*}
Together with  (\ref{AK8C}) this implies
\begin{eqnarray*}
e_{n}^\q(T_k,B_{C^r(Q_2)})&\le& c\left(\frac{n}{\log n\log \log (n+1)}\right)^{-\left(\frac{r}{d}+1\right)}(\log n)^5\\
&\le& cn^{-\left(\frac{r}{d}+1\right)}(\log n)^{\frac{r}{d}+6+\epsilon_0}.
\end{eqnarray*}
Finally, if $r+d+\s>(r+d)\frac{d_1}{d}$, 
we choose $\delta_1=0$ and $\delta_2>0$ so that 
$$
r+d+\s-\delta_2\left(r+d\right)>\left(r+d\right) \frac{d_1}{d}.
$$
From (\ref{AK9}),
\begin{eqnarray*}
e_{\bar{n}}^\q(T_k,B_{C^r(Q_2)})&\le&  c m^4 2^{-\left(r+d\right) \frac{d_1}{d}m}
\le cn^{-\left(\frac{r}{d}+1\right)}(\log n)^{4},
\end{eqnarray*}
which together with (\ref{AK8B}) shows that
\begin{eqnarray*}
e_n^\q(T_k,B_{C^r(Q_2)})&\le& c\left(\frac{n}{\log\log (n+1)}\right)^{-\left(\frac{r}{d}+1\right)}(\log n )^4\\
&\le&cn^{-\left(\frac{r}{d}+1\right)}(\log n)^{4+\epsilon_0}.
\end{eqnarray*}
\end{proof}
\section{Elliptic PDE}
Let $d,m\in \N$, $d\ge 2$, let $Q\subset \R^d$ be a $C^\infty$ domain (see, e.g., \cite{Hei05b} for the definition), and
let $\mathscr{L}$ be an elliptic differential operator of order $2m$ on $Q$, that is
\begin{equation}
\label{SO1}
\mathscr{L}u=\sum_{|\alpha|\le 2m} a_\alpha(x) D^\alpha u(x),
\end{equation}
with boundary operators
\begin{equation}
\mathscr{B}_ju=\sum_{|\alpha|\le m_j} b_{j\alpha}(x)D^\alpha u(x),
\end{equation}
where $j=1,\dots, m$, $m_j\le 2m-1$ and
$a_\alpha\in C^\infty(Q)$ and $b_{j\alpha}\in C^\infty(\partial Q)$ are complex-valued infinitely differentiable functions.
 We study
the homogeneous boundary value problem
\begin{eqnarray}
\mathscr{L}u(x)&=&f(x)\quad (x\in Q^0)\label{A5}\\
\mathscr{B}_ju(x)&=&0\quad (x\in \partial Q).\label{A6}
\end{eqnarray}
Let
\begin{eqnarray*}
a(x,\xi)&:=&\sum_{|\alpha|= 2m}a_\alpha(x) \xi^\alpha   \quad(x\in Q,\;\xi\in \R^d)\\
b_j(x,\xi)&:=&\sum_{|\alpha|= m_j}b_{j\alpha}(x) \xi^\alpha \quad(x\in \partial Q,\;\xi\in \R^d,\; j=1,\dots, m).
\end{eqnarray*}
We assume the ellipticity condition: 
$$
a(x,\xi)\ne 0 \quad (x\in Q,\; \xi \in \R^d\setminus\{0\})
$$
and for all linearly independent  $\xi,\eta\in \R^d$ the polynomial $a(x,\xi+\tau \eta)$ 
has exactly $m$ roots $\tau_i^+ \;(i=1,\dots,m)$ with positive imaginary part. Put
$$
a^+(x,\xi,\eta,\tau)=\prod_{i=1}^m(\tau-\tau_i^+).
$$
We also assume the complementarity condition: For all $x\in \partial Q$ and all
$\xi_x,\nu_x\in \R^d\setminus\{0\}$, where $\xi_x$ is tangent to $\partial Q$ at $x$ and $\nu_x$ is
orthogonal to the tangent hyperplane at $x$, the set of polynomials 
$ b_j(x,\xi_x+\tau\nu_x)\; (j=1,\dots, m)$ is linearly independent modulo $a^+(x,\xi_x,\nu_x,\tau)$.
Finally we suppose that there is a $\kappa_0$ with $0<\kappa_0<1$ such 
that for all $f$ in the H\"older space
$C^{\kappa_0}(Q)$ the classical solution $u$
exists and is unique (see  \cite{Kra67} and also, e.g., \cite{ADN64}, for the assumptions made here). 

Let $M$ be a smooth submanifold of $Q$ of dimension $d_1$, where $0\le d_1\le d$ 
(see, again, \cite{Hei05b} for a definition).
If $d_1=0$, we assume $M=\{x\}$, where $x$ is any inner point of $Q$. Let $r\in\N$, 
$F=B_{C^r(Q)}$, $G=L_\infty(M)$, 
and let $S:F\to G$ be given as
\begin{equation}
\label{CC2}
Sf=u|_{M},
\end{equation}
where $u$ is the solution of (\ref{A5}), (\ref{A6}). 
So we want to  find an approximation of the solution of  (\ref{A5}), (\ref{A6}) on a $d_1$-dimensional
submanifold $M$ of the domain $Q$, for right-hand sides belonging to $B_{C^r(Q)}$, and the error is measured in the
$L_\infty(M)$ norm.
We  put $K=\C$ and
\begin{equation}
\label{CC3}
\Lambda=\{\delta^\alpha_x\,:\,x\in Q, \: |\alpha | \le r\},
\end{equation}
where $\delta^\alpha_x(f)=D^\alpha f(x)$, that is, we 
allow information consisting of values of $f$ and its derivatives up to order $r$.

\begin{theorem}\label{theo:1} There are constants $c_1,c_2>0$ such that for all $n\in \N$ with $n\ge 2$,

\begin{eqnarray}
c_1n^{-\min\left(\frac{r+2m}{d_1},\,\frac{r}{d}+1\right)}&\le& e_n^\q(S,F)\nonumber\\
&\le & c_2 n^{-\min\left(\frac{r+2m}{d_1},\,\frac{r}{d}+1\right)}(\log n)^{\kappa(2m-d)},
\label{C2}
\end{eqnarray}
with $\kappa$ as defined in (\ref{AL1}) and (\ref{AL6}).
\end{theorem}

\begin{proof} By a result of Krasovskij \cite{Kra67}, Theorem 3.3 and Corollary, there is a 
kernel  $k\in\mathcal{C}^{\infty,2m-d}(Q,Q)$ such that for all $f\in C^{\kappa_0}(Q)$ the solution $u$ of 
(\ref{A5}), (\ref{A6}) 
satisfies
\begin{equation}
\label{CB1}
u(x)=\int_Q k(x,y)f(y)dy \quad (x\in Q).
\end{equation}
Consequently
$$
(Sf)(x)=(T_kf)(x) \quad (x\in M),
$$
that means, $S=T_k$, with $T_k$ considered as an operator from $C^r(Q)$ to $L_\infty(M)$. 
First we prove the upper bound. We show that it
holds  even for the smaller sets of information functionals
$\Lambda=\{\delta_x:\, x\in Q\}$. 
Let $Q_1=[0,1]^{d_1}$, considered as a subset of $\R^{d}$ by identifying $\R^{d_1}$ with $\R^{d_1}\times \{0^{(d-d_1)}\}$,
and let $Q_2=[0,1]^{d}$. The following representation of $T_k$ was shown in \cite{Hei05b}, proof of the upper bound in Theorem 1: 
There is
a $p\in\N$ (depending only on $M$ and $Q$) such that 
\begin{equation}
\label{SO4}
T_k=\sum_{i=1}^p \left(\bar{X}_iT_{k_i}Y_i+\bar{X}_iT_{h_i}J\right).
\end{equation}
Here 
\begin{eqnarray*}
\bar{X}_i&:&L_\infty(Q_1)\to L_\infty(M)\\
Y_i&:&C^r(Q)\to C^r(Q_2)
\end{eqnarray*}
$(i=1,\dots,p)$ are bounded linear operators and $J:C^r(Q)\to C(Q)$ is the embedding. Moreover, 
$Y_i$ is of the form (\ref{XG3}) with $\Lambda=\{\delta_x:\, x\in Q\}$, $\wt{\Lambda}=\{\delta_{x}:\, x\in Q_2\}$, and $\kappa=1$.
The kernels satisfy
\begin{eqnarray}
k_i&\in& \mathcal{C}^{\infty,2m-d}(Q_1,Q_2),\label{B1}\\
h_i&\in& \bigcap_{\s>0}\mathcal{C}^{\infty,\s}(Q_1,Q),\label{B2}
\end{eqnarray}
the integral operator $T_{k_i}$ is considered as acting from  $C^r(Q_2)$ to $L_\infty(Q_1)$, 
and $T_{h_i}$ is considered as a mapping from 
$C(Q)$ to $L_\infty(Q_1)$. (Using the terminology of \cite{Hei05b}: up to shifting and scaling of cubes, $\bar{X}_i$ stands for 
the product $E_iX_i$, $k_i$ is 
 the $k_i'$ from relation (64) of \cite{Hei05b}, and $h_i$ is $k_i''$ from relation (66)
of that paper, extended by zero to all of $Q$.)

Let $\nu_0,\nu_1$ be the smallest natural numbers satisfying 
$$
pe^{-\nu_0/8}\le 1/8,\quad e^{-\nu_1/8}\le 1/8,
$$
respectively. Let $c_1=\nu_0(3\nu_1+2)p$. Then, by Proposition \ref{pro:3},
\begin{eqnarray*}
\lefteqn{e_{c_1n}^\q(T_k,B_{C^r(Q)})   }\\
&\le& 2\sum_{i=1}^p \left(e_{2n}^\q(\bar{X}_iT_{k_i}Y_i,B_{C^r(Q)})+e_{3\nu_1 n}^\q(\bar{X}_iT_{h_i}J,B_{C^r(Q)}) \right).
\end{eqnarray*}
By Proposition \ref{pro:1}, 
\begin{eqnarray*}
e_{2n}^\q(\bar{X}_iT_{k_i}Y_i,B_{C^r(Q)})
&\le& \|\bar{X}_i\|e_{n}^\q(T_{k_i},\|Y_i\|B_{C^r(Q_2)})\\
&=& \|\bar{X}_i\| \|Y_i\|e_{n}^\q(T_{k_i},B_{C^r(Q_2)}).
\end{eqnarray*}
Furthermore, by Proposition \ref{pro:5},
\begin{eqnarray*}
\lefteqn{e_{3\nu_1 n}^\q(\bar{X}_iT_{h_i}J,B_{C^r(Q)}) }\\
&\le& 8 e_{n}^\de(J,B_{C^r(Q)})\sup_{\mathcal{E}\in\mathscr{C}(B_{C(Q)})}e_{n}^\q (\bar{X}_iT_{h_i},\mathcal{E}).
\end{eqnarray*}
Moreover, Lemma 1 of \cite{Hei04a} gives
$$
\sup_{\mathcal{E}\in\mathscr{C}(B_{C(Q)})}e_{n}^\q (\bar{X}_iT_{h_i},\mathcal{E})
\le \|\bar{X}_i\|\sup_{\mathcal{E}\in\mathscr{C}(B_{C(Q)})}e_{n}^\q (T_{h_i},\mathcal{E}).
$$
Thus we obtain
\begin{eqnarray}
\lefteqn{ e_{c_1n}^\q(T_k,B_{C^r(Q)})  }\nonumber\\
&\le& c\sum_{i=1}^p \Big(e_{n}^\q(T_{k_i},B_{C^r(Q_2)})\nonumber\\
&& +e_{n}^\de(J,B_{C^r(Q)})\sup_{\mathcal{E}\in\mathscr{C}(B_{C(Q)})}e_{n}^\q(T_{h_i},\mathcal{E})\Big).\label{AL2}
\end{eqnarray}
We conclude from (\ref{B1}) and Proposition \ref{pro:3a} that
\begin{equation}
\label{AL3}
e_{n}^\q(T_{k_i}, B_{C^r(Q_2)})\le cn^{-\min\left(\frac{r+2m}{d_1},\,\frac{r}{d}+1\right)}(\log n)^{\kappa(2m-d)},
\end{equation}
where $\kappa(2m-d) $ is  defined in (\ref{AL1}). Moreover, 
\begin{equation}
\label{AL4}
e_{n}^\de(J,B_{C^r(Q)})\le cn^{-\frac{r}{d}},
\end{equation}
see, e.g., \cite{Wen01}. Furthermore, from (\ref{B2}) and Proposition \ref{pro:2},
\begin{equation}
\label{AL5}
\sup_{\mathcal{E}\in\mathscr{C}(B_{C(Q)})}e_{n}^\q(T_{h_i},\mathcal{E})\le cn^{-1}(\log n)^{\alpha_1}
\end{equation}
where 
\begin{equation}
\label{QP1}
\a_1=\left\{\begin{array}{lll}
 0  & \mbox{if} \quad d_1<d,  \\
 4  & \mbox{if} \quad d_1=d. \\
    \end{array}
\right. 
\end{equation}
Relations (\ref{AL2})--(\ref{QP1}) finally give
\begin{equation}
\label{QP2}
 e_{c_1n}^\q(T_k,B_{C^r(Q)})\le cn^{-\min\left(\frac{r+2m}{d_1},\,\frac{r}{d}+1\right)}(\log n)^{\kappa(2m-d)}.
\end{equation}
Indeed, in the case $d_1<d$ this is clear. For $d_1=d$ we argue as follows: If $d+(2m-d)=2m\ge d=d_1$, then by (\ref{AL1}),
$\kappa(2m-d)=4$, which gives (\ref{QP2}). If $2m<d=d_1$, then $\frac{r+2m}{d_1}=\frac{r+2m}{d}<\frac{r}{d}+1$, hence
$$
n^{-\left(\frac{r}{d}+1\right)}(\log n)^4\le c n^{-\frac{r+2m}{d_1}},
$$
which leads to  (\ref{QP2}), again.
Now the desired upper bound in (\ref{C2}) follows from (\ref{QP2}) by rescaling. 

Next we prove the lower bound. As above, let $Q_1=[0,1]^{d_1}$, $Q_2=[0,1]^{d}$.
Let $C_0^r(Q_2)$ denote the subspace of 
$C^r(Q_2))$ consisting of those functions whose partial derivatives up to the order $r$ vanish on the boundary of $Q_2$.
It was shown in \cite{Hei05b}, section 5, that there are bounded linear operators
$X_0:C_0^r(Q_2)\to C^r(Q)$ and $Y_0:L_\infty(M)\to \C$ such that
\begin{equation}
\label{AM4}
Y_0SX_0=S_1,
\end{equation}
with $S_1:C_0^r(Q_2)\to \C$ the integration operator
$$
S_1f=\int_{Q_2}f(y)\,dy\quad (f\in C_0^r(Q_2)),
$$
and $X_0$ is of the form (\ref{XG3}) with $\Lambda=\{\delta_x^\a:\, x\in Q_2^0,\,|\a|\le r\}$,
$\wt{\Lambda}=\{\delta_x^\a:\, x\in Q,\,|\a|\le r\}$, and 
$\kappa$ depending only on $d$ and $r$. 
Consequently, by Proposition \ref{pro:1} and (\ref{AM4}),
\begin{eqnarray*}
e_{2\kappa n}^\q(S_1,B_{C_0^r(Q_2)})&\le&\|Y_0\| e_n^\q(S, \|X_0\|B_{C^r(Q)})\\
&=&\|X_0\|\|Y_0\| e_n^\q(S,B_{C^r(Q)}).
\end{eqnarray*}
From \cite{Nov01} it is known that
\begin{eqnarray*}
e_{2\kappa n}^\q(S_1,B_{C_0^r(Q_2)})&\ge& c n^{-\frac{r}{d}-1}.
\end{eqnarray*}
Thus we conclude 
\begin{eqnarray*}
e_n^\q(S,B_{C^r(Q)})&\ge& c n^{-\frac{r}{d}-1}.
\end{eqnarray*}
This proves the lower bound of (\ref{C2}) for the case $\frac{r}{d}+1\le \frac{r+2m}{d_1}$ (including the case $d_1=0$).

Now we assume $d_1\ge 1$. We use another reduction from \cite{Hei05b}, section 5, 
which will give the remaining part of the lower bound: There are bounded linear operators
$\bar{X}: C_0^{r+2m}(Q_1)\to C^r(Q)$  (representing the composition $\mathscr{L}XE$ from \cite{Hei05b})
and $Y:L_\infty(M)\to L_\infty(Q_1)$ such that 
\begin{equation}
\label{AM2}
YS\bar{X}=J,
\end{equation}
where $J:C_0^{r+2m}(Q_1)\to  L_\infty(Q_1)$ is the identical embedding, and $\bar{X}$ is of the form 
(\ref{XG3}) with $\Lambda=\{\delta^\alpha_x\,:\,x\in Q_1^0, \: \alpha\in \N_0^{d_1},\,|\alpha | \le r+2m\}$,
$\wt{\Lambda}=\{\delta^\alpha_x\,:\,x\in Q, \: \alpha\in \N_0^{d},\,|\alpha | \le r\}$, and $\kappa$ depending only on $d_1$, $m$ and $r$.
From Proposition \ref{pro:1} we obtain
$$
e_{2\kappa n}^\q\big(YS\bar{X},B_{C_0^{r+2m}(Q_1)}\big)\le \|Y\|e_n^\q\big(S, \|\bar{X}\|B_{C^{r}(Q)}\big)
\le c e_n^\q\big(S, B_{C^{r}(Q)}\big).
$$
Together with (\ref{AM2})
this yields,
$$
e_{2\kappa n}^\q\big(J,B_{C_0^{r+2m}(Q_1)}\big)\le c e_n^\q\big(S, B_{C^{r}(Q)}\big). 
$$
By \cite{Hei04b}, 
$$
e_{2\kappa n}^\q\big(J,B_{C_0^{r+2m}(Q_1)}\big)\ge cn^{-(r+2m)/d_1}.
$$
Consequently,
$$
e_{n}^\q\big(S, B_{C^{r}(Q)}\big)\ge cn^{-(r+2m)/d_1},
$$
concluding the proof of the lower bounds.
\end{proof}

\section{Comments}
In this section we recall previous results on the complexity of elliptic equations in the classical deterministic and randomized
setting and compare them with the results of the present paper. We discuss the speedups between the different settings.

Below $S$ and $F$ refer to the elliptic problem studied in section 6, see (\ref{CC2}). 
Let $e_n^\de(S,F)$ and $e_n^\ran(S,F)$ be the $n$-th minimal deterministic and randomized errors, respectively, 
as introduced, e.g., in \cite{Hei05b}, section 3. 

To suppress logarithmic factors, we use the following notation: For functions $a,b:\N_0\to [0,\infty)$ we write 
$a(n)\asymp_{\log} b(n)$ if there are constants 
$c_1,c_2>0$, $n_0\in \N_0$, $\alpha_1,\alpha_2\in \R$ such that 
$$
c_1(\log(n+2))^{\alpha_1}b(n)\le a(n)\le c_2(\log(n+2))^{\alpha_2}b(n)
$$
for all 
$n\in \N_0$ with $n\ge n_0$. Furthermore, we write $a(n)\asymp b(n)$ if the above holds with $\alpha_1=\alpha_2=0$. 

In the classical deterministic setting
we have
$$
e_n^\de(S,F)\asymp n^{-\frac{r}{d}}.
$$
This result is essentially contained in \cite{Wer91, Wer96, DNS05}, see also \cite{Hei05b}, where a proof is given for the specific 
function spaces considered here. Observe that in the deterministic setting the rate does not depend  on $d_1$, thus 
local and global problem are (up to constants) equally difficult.
As established in \cite{Hei05b}, in the classical randomized setting
we have
$$
e_n^\ran(S,F)\asymp_{\log} n^{-\min\left(\frac{r+2m}{d_1},\,\frac{r}{d}+\frac{1}{2}\right)}.
$$
By Theorem \ref{theo:1}, in the quantum setting,
$$
e_n^\q(S,F)\asymp_{\log}
n^{-\min\left(\frac{r+2m}{d_1},\,\frac{r}{d}+1\right)}.
$$
Thus, as in the classical randomized setting, the rate in the quantum setting depends on $d_1$. Note that the rate is  
$n^{-\frac{r}{d}-1}$ (the same as
that of quantum integration of functions from $C^r(Q)$, see \cite{Nov01}), for all $d_1\le\min(d,2m)$ and $r\in \N$.
Indeed, if $d\le 2m$, we infer
$$
\frac{r+2m}{d_1} \ge \frac{r+2m}{d} \ge \frac{r}{d}+1,
$$
while, if $d>2m$ and $d_1\le 2m$, we have
$$
\frac{r+2m}{d_1} \ge \frac{r+2m}{2m} > \frac{r}{d}+1.
$$
Let us compare the quantum setting with the classical deterministic
setting. We have a speedup for all $0\le d_1\le d$: For example, for all 
$d_1\le\min(d,2m)$, the speedup is $n^{-1}$. Furthermore, if $d_1=d$ and 
$d>2m$, the speedup is still $n^{-\frac{2m}{d}}$.

Comparing the quantum  with the classical randomized setting, we see that
for $d_1\le \min(d,2m)$
there is a speedup of $n^{-\frac{1}{2}}$, while for $d_1=d$ and $2m <d< 4m$ the speedup is $n^{\frac{1}{2}-\frac{2m}{d}}$, and
there is no speedup at all for $d_1=d$ and $d \ge 4m$. 

\end{document}